\documentclass{emulateapj}
\usepackage{times,mathptm}
\usepackage{lscape}

\newcommand {\asec} {$^{\prime\prime}$}

\newcommand{\Lsolar}{\mbox{\,$\rm L_{\odot}$}}
\newcommand{\Msolar}{\mbox{\,$\rm M_{\odot}$}}

\def\amin{\ifmmode ^{\prime}\else$^{\prime}$\fi}
\def\asec{\ifmmode ^{\prime\prime}\else$^{\prime\prime}$\fi}

\def\etal{{et\,al.\,}}

\def\chandra{{\it Chandra\/}}
\def\conx{{\it Constellation-X\/}}

\def\heao1{{\it HEAO-1\/}}

\def\iso{{\it ISO\/}}

\def\spitzer{{\it Spitzer\/}}
\def\xeus{{\it XEUS\/}}

\def\scuba{{\rm SCUBA\/}}

\def\ltsima{$\; \buildrel < \over \sim \;$}
\def\simlt{\lower.5ex\hbox{\ltsima}}
\def\gtsima{$\; \buildrel > \over \sim \;$}
\def\simgt{\lower.5ex\hbox{\gtsima}}

\journalinfo{The Astrophysical Journal, 2005 October, astro-ph/0506608}

\begin{document}
%

%

\title{The X-ray Spectral Properties of SCUBA Galaxies}

%

\author{D.M.~Alexander,\altaffilmark{1} F.E.~Bauer,\altaffilmark{1}
  S.C.~Chapman,\altaffilmark{2} I.~Smail,\altaffilmark{3}
  A.W.~Blain,\altaffilmark{2} W.N.~Brandt,\altaffilmark{4} and
  R.J.~Ivison\altaffilmark{5,6}}

\affil{$^1$Institute of Astronomy, Madingley Road, Cambridge CB3 0HA, UK}

\affil{$^2$California Institute of Technology, Pasadena, CA 91125, USA}

\affil{$^3$Institute for Computational Cosmology, University of Durham, South Road, Durham DH1 3LE, UK}

\affil{$^4$Department of Astronomy and Astrophysics, Pennsylvania State University, 525 Davey Laboratory, University Park, PA 16802, USA}

\affil{$^5$Astronomy Technology Centre, Royal Observatory, Blackford Hill, Edinburgh EH9 3HJ, UK}

\affil{$^6$Institute for Astronomy, University of Edinburgh, Blackford Hill, Edinburgh EH9 3HJ, UK}


\shorttitle{THE X-RAY SPECTRAL PROPERTIES OF SCUBA GALAXIES}

\shortauthors{ALEXANDER ET AL.}

\slugcomment{Received 2005 Feb 22; accepted 2005 June 23}

%
\begin{abstract}
%
    
  Deep \scuba\ surveys have uncovered a large population of massive
  submillimeter emitting galaxies (SMGs; $f_{\rm 850\mu
    m}\simgt$~4~mJy) at $z\simgt1$. Although it is generally believed
  that these galaxies host intense star-formation activity, there is
  growing evidence that a substantial fraction also harbor an Active
  Galactic Nucleus [AGN; i.e., an accreting super-massive black hole
  (SMBH)]. We present here possibly the strongest evidence for this
  viewpoint to date: the combination of ultra-deep X-ray observations
  (the 2~Ms \chandra\ Deep Field-North) and deep Keck spectroscopic
  data of SMGs with radio counterparts. We find that the majority
  ($\approx$~75\%) of these radio-selected spectroscopically
  identified SMGs host AGN activity; the other $\approx$~25\% have
  X-ray properties consistent with star formation (X-ray derived
  star-formation rates of $\approx$~1300--2700~\Msolar yr$^{-1}$). The
  AGNs have properties generally consistent with those of nearby
  luminous AGNs ($\Gamma\approx1.8\pm0.5$, $N_{\rm
    H}\approx$~$10^{20}$--$10^{24}$~cm$^{-2}$, and $L_{\rm
    X}\approx$~$10^{43}$--$10^{44.5}$~erg~s$^{-1}$) and the majority
  ($\approx$~80\%) are heavily obscured ($N_{\rm
    H}\simgt10^{23}$~cm$^{-2}$).  We construct composite rest-frame
  2--20~keV spectra for three different obscuration classes ($N_{\rm
    H}<10^{23}$~cm$^{-2}$, $N_{\rm
    H}=$~1--5~$\times10^{23}$~cm$^{-2}$, and $N_{\rm
    H}>5\times10^{23}$~cm$^{-2}$) which reveal features not seen in
  the individual X-ray spectra. An $\approx$~1~keV equivalent width
  Fe~K$\alpha$ emission line is seen in the composite X-ray spectrum
  of the most heavily obscured AGNs, suggesting Compton-thick or near
  Compton-thick absorption. Even taking into account the effects of
  absorption, we find that the average X-ray to far-infrared
  luminosity ratio of the AGN-classified SMGs (${L_{\rm
      X}}\over{L_{\rm FIR}}$=~0.004) is approximately one order of
  magnitude below that found for typical quasars. This result suggests
  that intense star-formation activity (of order
  $\approx$~1000~\Msolar yr$^{-1}$) dominates the bolometric output of
  these SMGs. However, we also explore the possibility that the X-ray
  to far-infrared luminosity ratio of the AGN components is
  intrinsically less than that found for typical quasars and postulate
  that some SMGs may be AGN dominated.  We investigate the
  implications of our results for the growth of massive black holes,
  discuss the prospects for deeper X-ray observations, and explore the
  scientific potential offered by the next generation of X-ray
  observatories.

\end{abstract}


%
\section{Introduction}\label{intro}
%

Deep \scuba\ surveys have uncovered a large population of
submillimeter (submm; $\lambda$~=~300--1000~$\mu$m) emitting galaxies
(SMGs; $\simlt$~1000 sources deg$^{-2}$ at $f_{\rm 850\mu
  m}\simgt$~\hbox{4~mJy}; e.g.,\ Smail, Ivison, \& Blain 1997; Barger,
Cowie \& Sanders 1999; Blain \etal 1999, 2002; Eales \etal 1999; Cowie
\etal 2002; Scott \etal 2002; Borys \etal 2003; Webb \etal 2003a). The
majority of these SMGs are faint at all other wavelengths, hindering
source identification studies (e.g.,\ Hughes \etal 1998; Ivison \etal
2000; Smail \etal 2002; Webb \etal 2003b). However, due to a
considerable amount of intensive multi-wavelength follow-up effort, it
is becoming clear that almost all are dust-enshrouded luminous
galaxies at \hbox{$z>1$} (e.g.,\ Ivison \etal 1998, 2000; Barger \etal
1999; Chapman \etal 2003a, 2005; Simpson \etal 2004; Pope \etal 2005).
With estimated bolometric luminosities of order $10^{13}$~\Lsolar\
(Blain \etal 2004a; Chapman \etal 2005), SMGs outnumber comparably
luminous galaxies in the local Universe by several orders of magnitude
(e.g.,\ Cowie \etal 2004). Dynamical mass and galaxy clustering
studies of spectroscopically identified SMGs have indicated that they
are likely to be massive galaxies (e.g.,\ Frayer \etal 1998, 1999;
Genzel \etal 2003; Neri \etal 2003; Blain \etal 2004b; Swinbank \etal
2004; Greve \etal 2005); these SMGs are $\approx$~5 times more massive
than the coeval optically selected field galaxies (e.g.,\ Erb \etal
2003; Steidel \etal 2004; Swinbank \etal 2004).  Making reasonable
assumptions about the duration of the submm-bright phase in these
systems, these results provide evidence that SMGs are the progenitors
of local $\simgt{M_{*}}$ early-type galaxies (e.g.,\ Lilly \etal 1999;
Scott \etal 2002; Smail \etal 2004; Chapman \etal 2005).

Central to the study of SMGs is the physical origin of their extreme
luminosities [i.e.,\ starburst or Active Galactic Nucleus (AGN)
activity]. If these sources are shown to be ultra-luminous starburst
galaxies then their derived star-formation rates suggest a substantial
increase in star-formation activity at $z>1$ (e.g.,\ Blain \etal 1999;
Smail \etal 2002). Conversely, if these sources are shown to be
powerful AGNs then they outnumber comparably luminous optical quasars
by $\approx$~1--2 orders of magnitude. The apparent association of a
few SMGs with quasars and their comparable comoving space densities
(when corrected for probable source lifetimes) suggest an evolutionary
connection between these two populations (e.g.,\ Page \etal 2001,
2004; Croom \etal 2004; Stevens \etal 2003, 2004; Chapman \etal 2005).
Given this evidence it is plausible that SMGs represent a rapid
black-hole and stellar growth stage prior to a luminous quasar phase
(e.g.,\ Archibald \etal 2002; Almaini 2003; Page \etal 2004; Alexander
\etal 2005a).

Arguably the most direct indication of AGN activity is the detection
of luminous hard X-ray emission (i.e.,\ $>$~2~keV). Hard X-ray
emission appears to be a universal property of AGNs, giving a direct
``window'' on the emission regions closest to the SMBH (e.g.,\
Mushotzky, Done, \& Pounds 1993), and it can provide a secure AGN
identification in sources where the optical signatures and
counterparts are weak or even non existent (e.g.,\ Alexander \etal
2001; Comastri \etal 2002). Hard X-ray emission is relatively
insensitive to obscuration (at least for sources that are Compton
thin; i.e.,\ $N_{\rm H}\simlt1.5\times10^{24}$~cm$^{-2}$) and any hard
X-ray emission from star formation in the host galaxy is often
insignificant when compared to that produced by the AGN. 

The first cross-correlation studies of moderately deep
(\hbox{$\approx$~20--200~ks}) \chandra\ surveys with \scuba\ surveys
yielded little overlap between the X-ray and submm detected source
populations ($\simlt$10--20\%; e.g.,\ Fabian \etal 2000; Bautz \etal
2000; Hornschemeier \etal 2000; Severgnini \etal
2000).\footnote{Further cross-correlation studies with moderately deep
  X-ray observations have revealed some overlap (e.g.,\ Barger \etal
  2001a; Ivison \etal 2002; Almaini \etal 2003; Waskett \etal 2003).
  This was probably due to the larger areal coverage of the \scuba\
  observations in these studies.}  To first order these results
suggested that bolometrically dominant AGNs can only be present in
typical SMGs if they are Compton thick. Later studies with the
\chandra\ Deep Field-North (CDF-N; Brandt \etal 2001, Alexander \etal
2003a) survey showed that a significant fraction (upwards of
$\approx$~30--50\% when the Borys \etal 2003 \scuba\ map is used) of
SMGs are X-ray detected (Barger \etal 2001b; Alexander \etal 2003b;
Borys \etal 2004). Direct X-ray spectral analyses of the five
AGN-classified SMGs in Alexander \etal (2003b) indicated that the AGNs
were heavily obscured but only moderately luminous at X-ray energies
($L_{\rm X}\approx$~0.3--1$\times10^{44}$~erg~s$^{-1}$ when corrected
for the effects of absorption). A comparison of the X-ray-to-submm
spectral slopes of these SMGs to that of NGC~6240 (a nearby luminous
galaxy with an obscured AGN) suggested that the AGNs typically
contributed only a few percent of the bolometric luminosity. However,
the small sample size and lack of spectroscopic redshifts (only one
source had a spectroscopic redshift) prevented more quantitative
conclusions.

The deep optical spectroscopic work of Chapman \etal (2003a, 2005) has
recently provided spectroscopic redshifts for 73 radio-identified
SMGs, a significant increase in sample size over previous studies
(e.g.,\ Ivison \etal 1998, 2000; Barger \etal 1999; Ledlow \etal 2002;
Smail \etal 2003; Simpson \etal 2004). The 2~Ms CDF-N field was one of
the regions targeted for this intensive spectroscopic follow up. The
combination of deep optical spectroscopic data and ultra-deep X-ray
observations provides powerful constraints on AGNs in SMGs. In
particular, spectroscopic redshifts improve the accuracy of the X-ray
spectral analyses over those of Alexander \etal (2003b) through the
identification of discrete X-ray spectral features (e.g.,\
Fe~K$\alpha$ emission) and the determination of any intrinsic
absorption, which is a strong function of redshift in a given X-ray
band [$N_{\rm H, z}\approx N_{\rm H, z0}(1+z)^{2.6}$].  These
improvements promise the most accurate determination of the AGN
contribution to the bolometric output of these SMGs to date. In this
paper we investigate the X-ray properties of the radio-selected
spectroscopically identified SMGs in the CDF-N field and predict the
AGN contribution to the bolometric luminosity of these SMGs. The
Galactic column density toward the CDF-N field is $(1.3\pm 0.4)\times
10^{20}$~cm$^{-2}$ (Lockman 2004), and
$H_{0}=65$~km~s$^{-1}$~Mpc$^{-1}$, $\Omega_{\rm M}=\onethird$, and
$\Omega_{\Lambda}=\twothirds$ are adopted throughout.

\vspace{0.5in}

%
\section{Observations and Basic Analysis}\label{props}
%

\subsection{\scuba\ Galaxy Sample}

Our \scuba\ galaxy sample includes 20 SMGs that have been
spectroscopically identified by Chapman \etal (2005) and lie in the
region of the 2~Ms CDF-N observation (Alexander et~al. 2003a). The
optical counterparts for these sources were chosen based on the
positional association with a radio source [identified by either {\rm
  VLA} (Richards 2000) or {\rm MERLIN} (Chapman \etal 2004a)
observations]. The basic properties of the SMGs are given in Table~1.
The source redshifts ($z=$~0.555--2.914 with a median of
$z=$~2.0$\pm$0.6) are typical of the spectroscopically identified SMG
population (median redshift of $z\approx$~2.2; Chapman \etal 2005).
The $\approx$~35--50\% of sources without radio counterparts (e.g.,\
Ivison \etal 2002; Chapman \etal 2003b; Borys \etal 2004) could not be
targeted with optical spectroscopy. However, the range of 850$\mu$m
flux densities for the submm sample ($f_{\rm 850\mu
  m}\approx$~4--12~mJy, with a mean of $6.6\pm2.2$~mJy) is typical of
SMGs detected in blank-field \scuba\ surveys (e.g.,\ Scott \etal 2002;
Webb \etal 2003a).

The potential selection effects introduced by the radio and
spectroscopic-identification of this sample, compared to a purely
submm-flux-limited sample of SMGs, are discussed at length in Chapman
\etal (2005) and Blain \etal (2004). The need for both a faint radio
counterpart and identifiable spectral features to aid in redshift
measurement may result in this sample having a higher incidence of AGN
activity than the SMG population as a whole; however, the
incompleteness in our surveys due to these two selection criteria can
be equally well explained by the likely temperature and redshift
distribution of the SMG population (Chapman et al.\ 2005).
Nevertheless, we caution the reader that the sample analysed here may
not be completely representative of the entire $f_{\rm 850\mu
  m}\simgt$~4~mJy SMG population.

Many of the SMGs in our sample were radio-selected sources
specifically targeted with SCUBA observations; see the table in
Alexander \etal (2005a) for the observation modes. While this could
potentially cause an additional AGN bias over a purely radio-detected
SMG sample, we do not find evidence for a strong bias in our sample
(see Alexander \etal 2005a for the results of a two-sided Fisher's
exact test).

\subsection{\chandra\ Counterparts to the \scuba\ Galaxy Sample}

The 2~Ms CDF-N observations were centered on the optical Hubble Deep
Field-North (HDF-N; Williams \etal 1996) region and cover
$\approx448$~arcmin$^2$ (Alexander et~al. 2003a). These observations
provide the deepest view of the Universe in the 0.5--8.0~keV band; the
aim-point sensitivities are
\hbox{$\approx7.1\times10^{-17}$}~erg~cm$^{-2}$~s$^{-1}$ at 0.5--8.0~keV
(full band), \hbox{$\approx2.5\times10^{-17}$}~erg~cm$^{-2}$~s$^{-1}$ at
0.5--2.0~keV (soft band), and
\hbox{$\approx1.4\times10^{-16}$}~erg~cm$^{-2}$~s$^{-1}$ at 2--8~keV (hard
band). At the median redshift of the SMGs, the observed 0.5--8.0~keV
band corresponds to rest-frame energies of 1.5--24~keV.  Such
high X-ray energies allow very high column densities to be penetrated
(e.g.,\ at 20~keV, $\simlt$~55\% of the direct X-ray emission is
absorbed with column densities of $N_{\rm H}\simlt10^{24}$~cm$^{-2}$;
see Appendix~B in Deluit \& Courvoisier 2003).

Using a 1.5\asec\ search radius, 16 of the 20 SMGs were found to have
\chandra\ counterparts in the main \chandra\ catalog of Alexander
\etal (2003a). Given the low surface density of SMGs, we also searched
for X-ray counterparts using the complete supplementary \chandra\
catalog of Alexander \etal (2003a) and found one further match; the
probability of this match being spurious is \hbox{$<1$\%} (see \S2.4
of Alexander et~al. 2003b). In total 17 ($85^{+15}_{-20}$\%) of the 20
SMGs in our sample have a \chandra\ counterpart (see
Table~1).\footnote{All errors are taken from Tables 1 and 2 of Gehrels
  (1986) and correspond to the $1\sigma$ level; these were calculated
  assuming Poisson statistics.}  We have calculated rest-frame
0.5--8.0~keV luminosities, and 1.4~GHz luminosity densities following
equations 1 and 2 in Alexander \etal (2003b), assuming an X-ray
spectral slope of $\Gamma=1.8$ and a radio spectral slope of
$\alpha=0.8$, respectively.  Far-infrared (far-IR;
$\lambda$~=~40--120~$\mu$m) luminosities were calculated from the
rest-frame 1.4~GHz luminosity densities under the assumption of the
radio-to-far-IR correlation with $q=2.35$ (e.g.,\ Helou, Soifer, \&
Rowan-Robinson 1985); the presence of an AGN component to the radio
emission will lead to an overprediction of the far-IR luminosity. See
Table~1.

The three X-ray undetected SMGs lie in sensitive regions of the CDF-N
field and have 3~$\sigma$ flux limits below the fluxes of the X-ray
detected SMGs; see Table~1.  Their redshifts ($z=$~2.0--2.1) are
consistent with the median redshift of the SMG sample. Stacking the
X-ray data of the individually undetected SMGs following the procedure
of Lehmer \etal (2005) yields marginally significant detections in the
soft and full bands ($2.5\sigma$, soft band; $2.7\sigma$, full band;
S.~Immler, private communication), suggesting that these sources lie
just below the individual source detection limit. The corresponding
average X-ray constraints are
$\approx9.0\times10^{-17}$~erg~cm$^{-2}$~s$^{-1}$ (full band),
$\approx1.7\times10^{-17}$~erg~cm$^{-2}$~s$^{-1}$ (soft band) and
$<1.6\times10^{-16}$~erg~cm$^{-2}$~s$^{-1}$ (hard band). The full-band
constraint corresponds to $L_{\rm
  0.5-8.0~keV}\approx2\times10^{42}$~erg~s$^{-1}$ at $z=2$ for
$\Gamma=1.8$ and is consistent with that expected from luminous star
formation (see \S3.2 and \S4.2). 

\subsection{X-ray Spectra: Extraction and Fitting}

X-ray spectra were generated for all of the SMGs with full-band fluxes
\hbox{$>3\times10^{-16}$}~erg~cm$^{-2}$~s$^{-1}$, which typically
corresponds to $>50$ X-ray counts. To account for the range of roll
angles and aim points in the 20 separate observations that comprise
the 2~Ms \hbox{CDF-N,} the X-ray spectra were generated using the ACIS
source extraction code ({\sc ACIS Extract}) described in Broos \etal
(2002).\footnote{{\sc ACIS Extract} is a part of the {\sc TARA}
  software package and can be accessed from
  http://www.astro.psu.edu/xray/docs/TARA/ae\_users\_guide.html.}
Briefly, for each source this code extracts the counts from each of
the observations taking into account the changing shape and size of
the PSF with off-axis angle as given in the \chandra\ X-ray Center
(CXC) PSF library.\footnote{See
  http://asc.harvard.edu/ciao2.2/documents\_dictionary.html\#psf.} A
local background is extracted after all of the sources are excluded
from the X-ray event file, and the spectra and response matrices are
summed using standard {\sc ftools} routines (Blackburn 1995). See
F.~E.~Bauer et~al. (in preparation) for further information and the
X-ray spectral properties of the sources detected in the \chandra\
deep fields.

Given the limited counting statistics of the X-ray sources, we
performed all of the X-ray spectral analyses using the
$C$-statistic (Cash 1979). One advantage of using the $C$-statistic is
that the data can be fitted without binning, making it ideally suited
to low-count sources (e.g.,\ Nousek \& Shue 1989). A disadvantage of
using the $C$-statistic is that it is not possible to determine
rigorously if one model provides a statistically preferable fit than
another; however, it is possible to inspect the fit residuals and
perform Monte-Carlo analyses to distinguish between different models.
Hence in our analyses we have focused on empirically motivated models
which have proven to provide a robust characterisation of the
properties of well-studied AGNs. We used the latest version of {\sc
  xspec} (v11.3.1; Arnaud 1996, 2002) for all of the model fitting,
which allows the $C$-statistic to be used on background-subtracted
data; we carried out several checks of the background-subtraction
method to verify that no spurious residual features were present in
the background-subtracted data. We fitted the X-ray spectra of the
SMGs using a power law model (with Galactic absorption) in the
rest-frame 2--10~keV and 5--20~keV bands; see Table~2. The constraints
on the five sources with $\simlt100$ counts are poor. However, our
analyses have been designed to maximize the useful constraints for
even the weakest X-ray sources.  The fit parameter uncertainties are
quoted at the 90\% confidence level for one parameter of interest
(Avni 1976). We provide details of our analysis techniques below but
defer the discussion of our results to \S3.

%
%
\begin{figure}
\centerline{\includegraphics[angle=0,width=9.0cm]{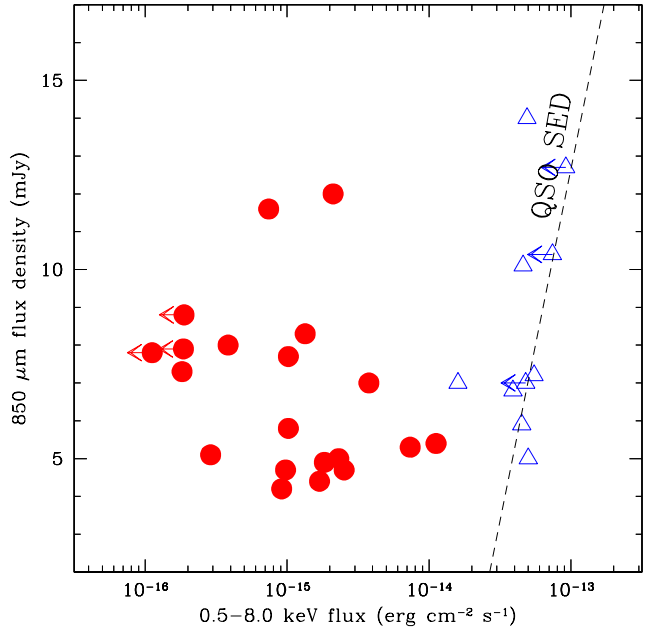}}
\vspace{-0.05in} 
\figcaption{Submm flux density versus full-band flux for the SMGs
  (filled circles) and optically classified quasars with X-ray and
  submm constraints (open triangles).  The quasar data are taken from
  Page \etal (2001), Vignali \etal (2001), and Isaak \etal (2002). The
  dashed line indicates the expected X-ray fluxes and submm flux
  densities for a quasar with the same properties as 3C273; the
  submm-to-X-ray spectral slope is independent of redshift (see
  Figure~2 of Fabian \etal 2000). Our spectroscopically identified
  SMGs are up to two orders of magnitude fainter in the X-ray band
  than the optically classified quasars for a given submm flux
  density.}  
\vspace{-0.15in}
\end{figure}

%
%
\begin{figure*}
\centerline{\includegraphics[angle=0,width=9.0cm]{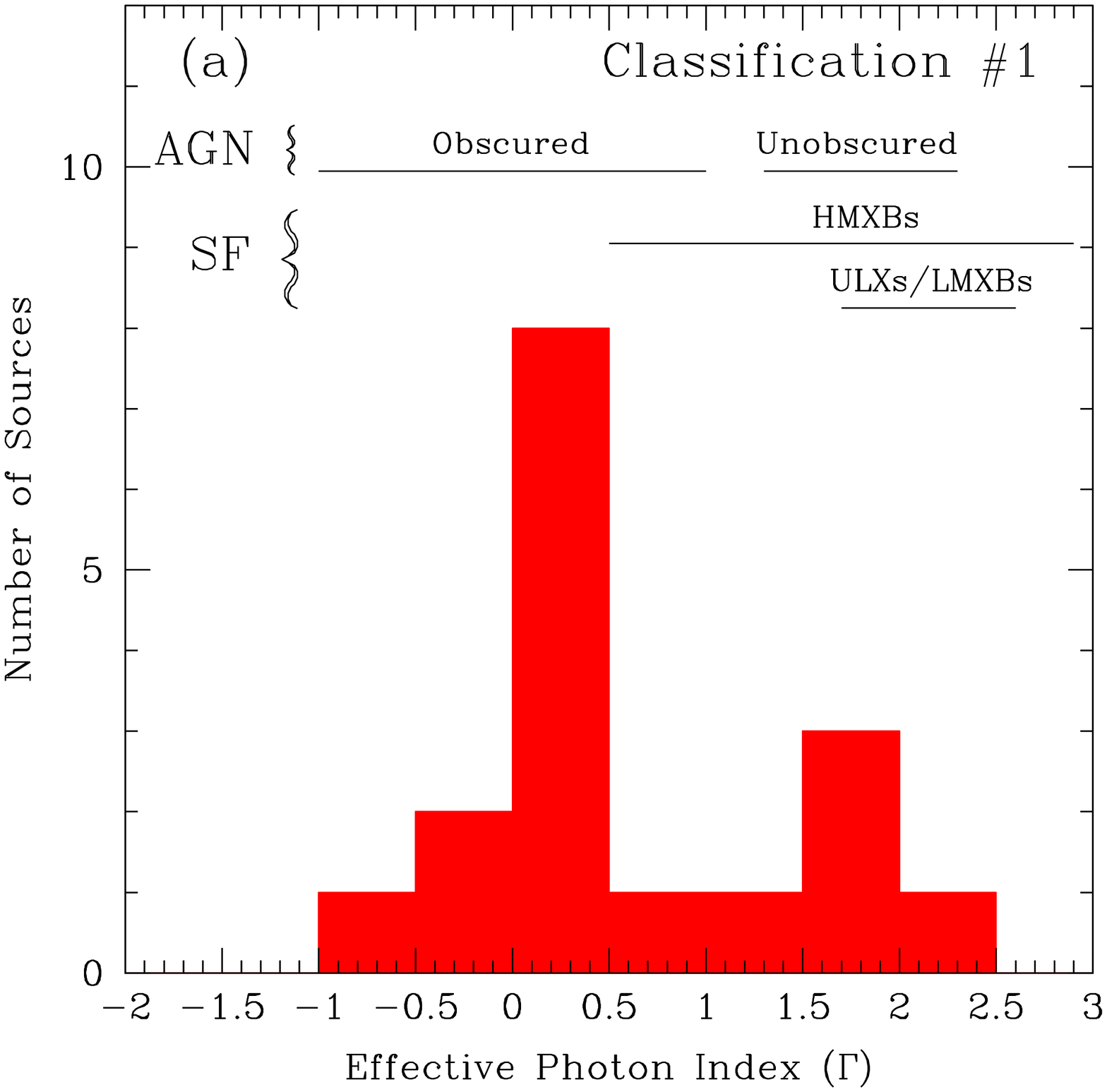}\hfill
\includegraphics[angle=0,width=9.0cm]{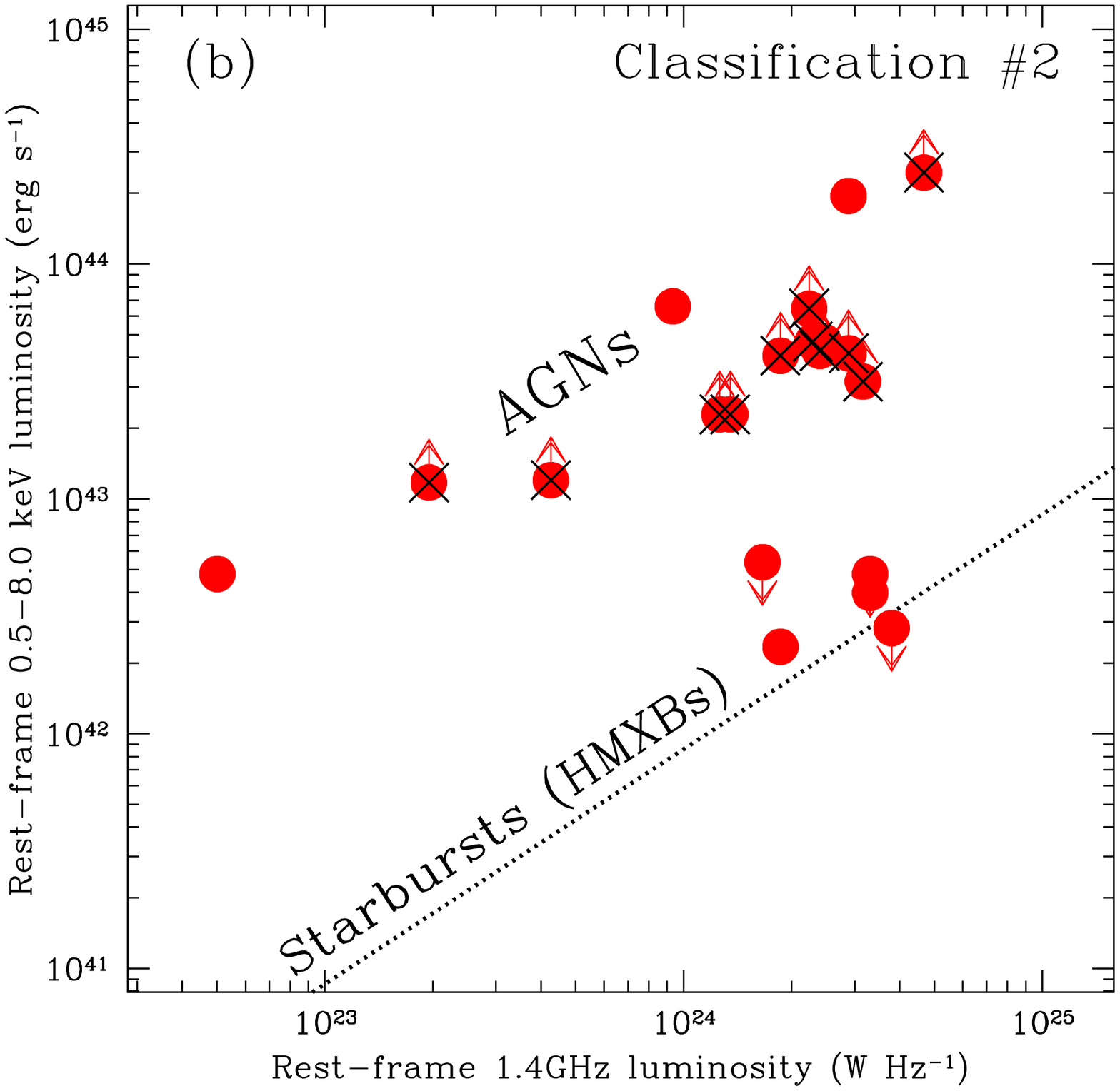}}
  \vspace{0.05in} \figcaption{Classification scheme for the X-ray
    emission from the SMGs. (a) Effective photon-index histogram. The
    approximate observed X-ray spectral slopes for a variety of
    different source types [split broadly into AGN and star formation
    (SF)] are shown (e.g.,\ Nandra \& Pounds 1994; Maiolino \etal
    1998; Colbert \etal 2004). The only sources that produce extremely
    flat or inverted X-ray spectral slopes ($\Gamma<0.5$) are obscured
    AGNs. The typical uncertainties in the X-ray spectral slopes for
    the plotted sources are $\Delta\Gamma\approx$~0.3; see Table~1.
    (b) Rest-frame 0.5--8.0~keV luminosity versus 1.4~GHz luminosity
    density. The sources classified as obscured AGNs from Figure~2a
    are indicated with crosses; the X-ray luminosities have not been
    corrected for the effect of absorption.  The dotted line shows the
    X-ray--radio relationship for star-forming galaxies whose X-ray
    emission is dominated by HMXBs (Persic \etal 2004); this
    relationship is converted to the 0.5--8.0~keV band from the
    2--10~keV band assuming $\Gamma=1.8$. The X-ray emission from 15
    ($\approx$~75\%) of the 20 SMGs is dominated by AGN activity. We
    caution the reader that due to selection biases this does not
    directly indicate a $\approx$~75\% AGN fraction in the bright SMG
    population; see \S3.2.}
\end{figure*}

In order to interpret the X-ray spectral fitting results we have
developed an empirically motivated AGN model and constructed a simple
rest-frame power-law diagnostic diagram. The advantage of this
approach is that we only need to perform simple rest-frame power law
fits to the sources, allowing us to determine the characteristics of
sources with few X-ray counts. Our adopted AGN model includes a
power-law component (which can be absorbed by neutral material;
$N_{\rm H}$), a neutral reflection component (the {\sc pexrav} model
in {\sc xspec}), a scattered component (Sc) of ionised gas, and a
Fe~K$\alpha$ emission line at 6.4~keV. The {\sc xpsec} model
components for our adopted model are {\sc (pow+pexrav+zga)+pow*zwabs}.

We used the {\sc xspec} command $fake\_it$ to produce the model tracks
for a range of X-ray spectral slopes ($\Gamma=$~1.3--2.3) and
absorbing column densities ($N_{\rm
  H}=$~$10^{20}$--2$\times10^{24}$~cm$^{-2}$).  The ratio of the
direct to reflected emission (i.e.,\ $rel\_refl$ in the {\sc pexrav}
model) was fixed to unity, in agreement with the average value found
for local AGNs (e.g.,\ Risaliti \etal 2002; Deluit \& Courvoisier
2003; Malizia \etal 2003). We set the other parameters of the neutral
reflection component to be the same as those in Model 2c of Malizia
\etal (2003).  The ionised gas scattering fraction was set to be 1\%
of the intrinsic power-law emission, and the equivalent width of the
Fe~K$\alpha$ emission line was varied depending on the absorbing
column density ($W_{\lambda}=0.145$~keV for $N_{\rm
  H}<3\times10^{23}$~cm$^{-2}$, $W_{\lambda}=0.37$~keV for $N_{\rm
  H}=$~3--7$\times10^{23}$~cm$^{-2}$, and $W_{\lambda}=1.0$~keV for
$N_{\rm H}>7\times10^{23}$~cm$^{-2}$; see Risaliti 2002 and Malizia
\etal 2003). We also investigated higher column densities with a
varying amount of ionised gas scattering fraction [$N_{\rm
  H}\approx10^{25}$~cm$^{-2}$ (i.e.,\ pure reflection) and Sc~=~0--10\%]. To
take account of the \chandra\ ACIS-I instrumental response we used the
response matrix file (RMF) and ancillary response file (ARF) from
CXOHDFN~J123629.1+621045. To allow for the construction of the model
tracks up to rest-frame energies of 20~keV we assumed a redshift of
$z=2.0$, consistent with the average redshift of the SMG sample.

We also attempted to fit directly the individual X-ray spectra of the
SMGs with the empirically motivated model outlined above. However, we
could not statistically constrain these models sufficiently to gain
additional physical insight over that found using the power-law
diagnostic diagram analyses. This was most likely due to a degeneracy
in the number of model components and the generally poor photon
statistics of the individual X-ray spectra. 

%
\section{X-ray Properties of the SCUBA Galaxy Sample}\label{data}
%

\subsection{Basic Source Properties}

We show the submm flux density versus full-band flux of all of the
SMGs in Figure~1. As a comparison we also show the submm and X-ray
properties of samples of moderate-to-high redshift quasars
($z\approx$~1--5; Page \etal 2001; Vignali \etal 2001; Isaak \etal
2002). The average redshift and submm flux density of the Page \etal
(2001) quasars ($z=2.1\pm0.4$ and $f_{\rm 850\mu m}=7.5\pm1.6$~mJy)
are consistent with those of our SMGs; however, the median X-ray flux
of the SMGs is $\approx$~30 times fainter. The submm properties of
quasars are proving to be important from the point of view of the
processes of star formation at high redshift and providing clues to
the relationship between the formation of SMGs, quasars, and massive
galaxies (e.g.,\ Stevens \etal 2003, 2004; Page \etal 2004). However,
the X-ray properties of spectroscopically identified SMGs can only be
investigated with ultra-deep X-ray observations.

\subsection{X-ray Source Classification}

Due to the exceptional sensitivity of the 2~Ms \hbox{CDF-N}
observations it is possible to detect X-ray emission from star
formation out to high redshift.  For example, the soft-band flux limit
at the aim point corresponds to $L_{\rm
  1.5-6~keV}>$~8~$\times10^{41}$~erg~s$^{-1}$ for a galaxy at $z=2$.
This luminosity is comparable to those for the most X-ray luminous
starburst galaxies in the local Universe (e.g.,\ NGC~3256; Moran \etal
1999; Lira \etal 2002), sources that are bolometrically less energetic
than SMGs by $\approx$~1 order of magnitude.  Therefore, the detection
of X-ray emission from an SMG does not {\it a priori} indicate the
presence of AGN activity. In order to classify the SMGs in our sample
we have used a two-tiered classification procedure. This
classification procedure is conservative and may not identify all of
the SMGs that host AGN activity. However, we can be confident that any
AGN classifications are secure.

The first method of source classification is based on the effective
X-ray spectral slope; see Figure~2a. Star-forming galaxies and
unobscured AGNs generally have comparatively steep X-ray spectral
slopes (i.e.,\ $\Gamma\approx$~2; e.g.,\ Kim, Fabbiano, \& Trinchieri
1992a,b; Nandra \& Pounds 1994; Ptak \etal 1999; George \etal 2000;
Colbert \etal 2004).\footnote{Detailed X-ray studies of local
  star-forming galaxies have shown that they have more complex X-ray
  spectra than simple power-law emission (e.g.,\ power-law emission
  and a very soft $\approx$~0.7~keV thermal component; Ptak \etal
  1999); however, at the probable redshifts of our sources the very
  soft thermal component will be redshifted out of the \chandra\
  energy band.}  Obscured AGNs are usually distinguishable from
unobscured AGNs and star-forming galaxies by the presence of a flat
X-ray spectral slope ($\Gamma<1$) due to the energy-dependent
photo-electric absorption of the X-ray emission (e.g.,\ Maiolino \etal
1998; Risaliti, Maiolino, \& Salvati 1999). However, since SMGs are
potentially massive galaxies undergoing intense star formation, they
could have a large population of high-mass X-ray binaries (HMXBs),
which may have comparatively flat X-ray spectral slopes (e.g.,\
neutron star HMXBs can have $\Gamma\approx$~0.5--1.0; Colbert \etal
2004). Eleven of the 17 X-ray detected SMGs have $\Gamma<0.5$,
unambiguously indicating the presence of a heavily obscured AGN.

The second method of source classification focuses on the expected
X-ray emission from star formation; see Figure~2b. Many studies have
shown that the radio luminosity can be used to predict the X-ray
luminosity in star-forming galaxies (e.g.,\ Shapley, Fabbiano, \&
Eskridge 2001; Bauer \etal 2002; Ranalli \etal 2003; Grimm, Gilfanov,
\& Sunyaev 2003; Gilfanov, Grimm, \& Sunyaev 2004; Persic \etal 2004).
The main assumption in these predictions is that the radio emission is
dominated by star formation; however, the presence of an AGN component
to the radio emission will only lead to an overprediction of the
contribution from star formation at X-ray energies. The normalisation
and slope of these X-ray--radio relationships differ depending upon
the relative contributions from low-mass and high-mass X-ray binaries
(LMXBs; HMXBs; Gilfanov \etal 2004; Persic \etal 2004; see Figure~2b).
Since the properties of SMGs indicate that they are undergoing intense
star-formation activity, we only need to consider the contribution
from HMXBs (see \S4.2). The predicted X-ray emission from star
formation in SMGs is large (up to $L_{\rm X}\approx
10^{42}$--$10^{43}$~erg~s$^{-1}$). Even so, all of the sources
classified as AGNs show an X-ray excess due to AGN activity; these
X-ray excesses will be even greater once the X-ray luminosities are
corrected for the effect of absorption. With this second method a
further four sources show a clear X-ray excess over that predicted
from star formation, indicating that their X-ray emission 
is dominated by AGN activity. Four of the 10 AGN-classified
SMGs with $>100$ counts also show some evidence for X-ray variability
(M.~Paolillo, private communication; see Table~1), a further signature
of AGN activity; these constraints are consistent with those found for
similarly bright AGNs in the \chandra\ Deep Field-South (Paolillo
\etal 2004). None of the sources that have X-ray emission consistent
with star formation show evidence for X-ray variability, although the
photon statistics are poor.

%
%
\begin{figure}
\centerline{\includegraphics[angle=0,width=9.0cm]{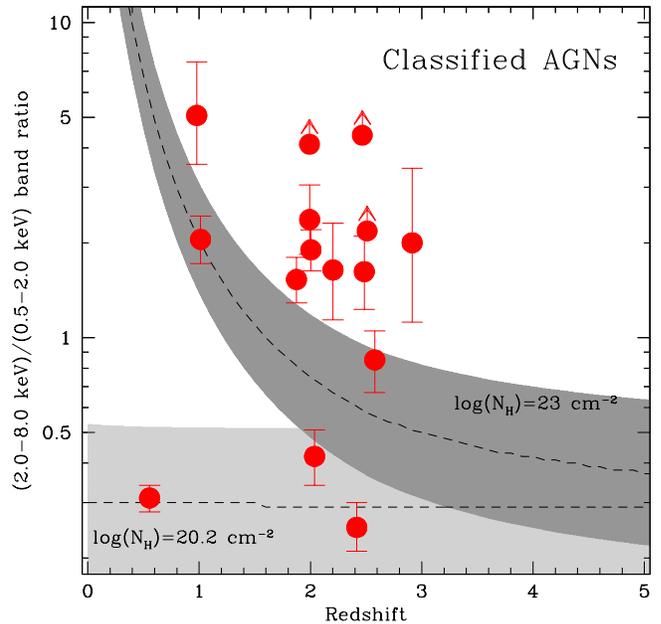}}
\vspace{-0.05in} 
\figcaption{X-ray band ratio versus spectroscopic
  redshift for the X-ray classified AGNs. The light and dark shaded
  regions show the range of expected band ratios for an unabsorbed and
  absorbed AGN, respectively. These regions were calculated assuming a
  $\Gamma=1.8\pm0.5$ power law with differing amounts of absorption
  (as shown); these simple AGN models have been calculated using {\sc
    pimms} Version 3.2d. The error bars correspond to the uncertainties in the
  band ratio. This simple figure has limited diagnostic utility;
  however, it suggests that almost all of the AGNs are obscured.} 
\end{figure}

Overall, the X-ray emission from 15 ($75^{+25}_{-19}$\%) of the 20
SMGs in the sample is AGN dominated.  The other five SMGs are the
faintest X-ray sources in the sample, and three sources are undetected
in the X-ray band (see \S2.2). The X-ray emission from these five
sources is likely to be dominated by star-formation activity (see
\S4.2 for further constraints); however, heavily obscured or X-ray
weak AGNs cannot be ruled out (e.g.,\ compare with the similar
$z=2.285$ galaxy FSC~10214+4724; Alexander \etal 2005b). We caution
the reader that, due to our sample selection and completeness, these
results do not directly indicate a $\approx$~75\% AGN fraction in the
bright SMG population (the AGN fraction is probably
$>38^{+12}_{-10}$\%; Alexander \etal 2005a, consistent with earlier
estimates from the X-ray data; e.g.,\ Alexander \etal 2003b; Borys
\etal 2004; Wang, Cowie, \& Barger 2004).

%
%
\begin{figure}
\centerline{\includegraphics[angle=0,width=9.0cm]{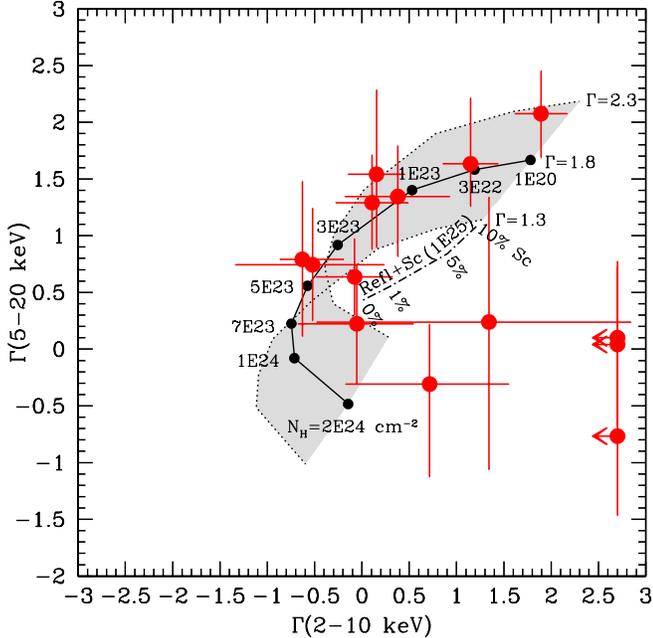}}
\vspace{0.05in}
\figcaption{Rest-frame 5--20~keV spectral slope versus rest-frame
  2--10~keV spectral slope showing the X-ray properties of individual
  SMGs. The shaded region indicates where typical AGNs with
  $\Gamma=1.8\pm0.5$ and $N_{\rm
    H}=10^{20}$--$2\times10^{24}$~cm$^{-2}$ are expected to lie
  (individual values of $N_{\rm H}$ and $\Gamma$ are indicated); see
  \S2.3. The dot-dashed line shows where completely Compton-thick AGNs
  (i.e.,\ $N_{\rm H}\approx10^{25}$~cm$^{-2}$; a pure reflection
  spectrum with ionised gas scattering) with different amounts of
  scattering (as indicated) are expected to lie; the assumed X-ray
  spectral slope is $\Gamma=1.8$. This figure shows that the
  AGN-classified SMGs are well represented by the model, implying
  that the variations in their X-ray spectral properties are due
  to absorption.} 
\end{figure}

%
\begin{figure}
\centerline{\includegraphics[angle=0,width=9.0cm]{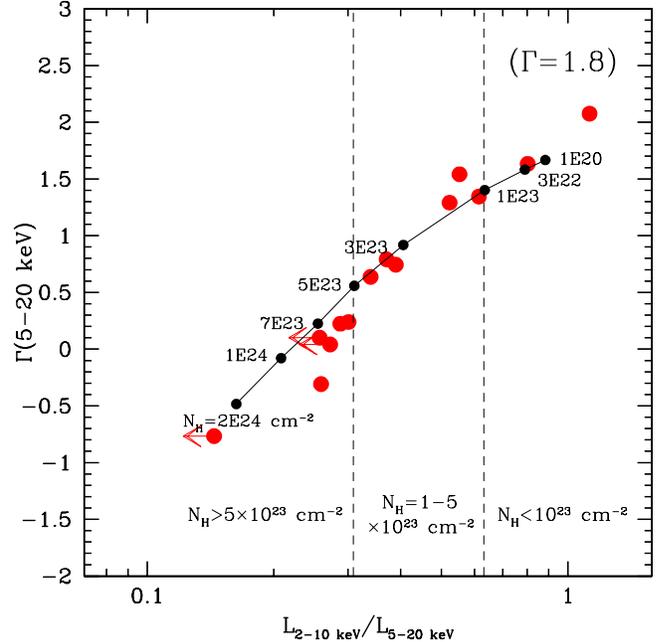}}
\vspace{0.05in} 
\figcaption{Rest-frame 5--20~keV spectral slope versus rest-frame
  X-ray luminosity ratio (${{L_{\rm 2-10 keV}}\over{L_{\rm 5-20
        keV}}}$). The solid line indicates the track for our adopted
  AGN model with $\Gamma=1.8$ (individual values of $N_{\rm H}$ are
  indicated). The dashed lines indicate the divisions of the three
  obscuration classes used in the subsequent analyses ($N_{\rm
    H}<10^{23}$~cm$^{-2}$, $N_{\rm H}=1-5\times10^{23}$~cm$^{-2}$, and
  $N_{\rm H}>5\times10^{23}$~cm$^{-2}$). The implied absorbing column
  densities of the AGNs in the SMGs are in good agreement with
  Figure~4.}  
\end{figure}

\subsection{X-ray Spectral Analyses}

The flat X-ray spectral slopes for the majority of the AGN-classified
SMGs indicates the presence of absorption. Basic X-ray absorption
constraints can be set using the X-ray band ratio (defined as the
ratio of the hard-band to soft-band count rate); see Figure~3.  This
approach has limited diagnostic utility because AGNs often have more
complex spectra than that of power-law emission with differing amounts
of absorption. However, it suggests that the majority of the sources
are heavily obscured (i.e.,\ $N_{\rm H}\simgt10^{23}$~cm$^{-2}$). The
intrinsic luminosity of the obscured AGNs could be considerably
greater than the observed luminosity if the obscuration is high.

In order to estimate the amount of absorption in each of the
AGN-classified SMGs more accurately we have constructed a rest-frame
power-law diagnostic diagram; see \S2.3 and Figure~4. The tracks on
this diagram show the approximate regions where typical AGNs are
likely to lie; since we investigate rest-frame energies up to 20~keV,
CXOHDFN~J123636.7+621156 with a redshift of only $z=0.555$ cannot be
included in these analyses. The AGN-classified SMGs are clearly well
represented by this diagram and, on the basis of our model tracks, 12
($\approx$~80\%) are heavily obscured ($N_{\rm
  H}\simgt10^{23}$~cm$^{-2}$). Under the assumption that the different
X-ray spectral properties of the SMGs are due to absorption, the lower
energy X-ray emission should be more strongly attenuated than the
higher energy X-ray emission. This is demonstrated in Figure~5 where
the implied attenuations from the luminosity ratios in the 2--10 and
5--20~keV bands are in agreement with those expected from the
rest-frame 5--20~keV spectral slopes and the results from Figure~4.

The constraints on the absorption properties of individual sources are
comparatively poor. To provide tighter overall constraints we
performed joint spectral fitting of all of the sources in the three
different obscuration classes indicated in Figure~5 ($N_{\rm
  H}<10^{23}$~cm$^{-2}$, $N_{\rm H}=$~1--5~$\times10^{23}$~cm$^{-2}$,
and $N_{\rm H}>5\times10^{23}$~cm$^{-2}$). We jointly fitted the
rest-frame 2--10~keV and 5--20~keV spectral slopes of the sources,
leaving the power-law normalisations as a free parameter for each
source. The results of this joint spectral fitting are shown in
Figure~6. The average absorbing column densities for the sources in
each obscuration class are in excellent agreement with those implied
from the individual X-ray spectral analyses. Furthermore, the average
intrinsic X-ray spectral slopes estimated from the model tracks
($\Gamma\approx$~1.8) are in good agreement with those found for
typical AGNs in the local Universe (e.g.,\ Nandra \& Pounds 1994;
George \etal 2000).

\subsection{Rest-frame 2--20~keV spectra}

The previous analyses have provided the broad-band X-ray spectral
properties of the AGN-classified SMGs but have not been sensitive to
discrete emission features (e.g.,\ Fe~K$\alpha$ emission). Unambiguous
Fe~K$\alpha$ emission has not been identified in any of the individual
X-ray spectra and, due to poor photon statistics, it is difficult to
constrain the X-ray continuum either side of the emission line, which
significantly affects the accuracy of Fe~K$\alpha$ constraints. It is
not possible to improve greatly the Fe~K$\alpha$ constraints via joint
spectral fitting since the critical parameters are the emission-line
and X-ray continuum normalisations, which vary from source to source.
However, we can improve the overall signal-to-noise ratio of the
spectra by stacking the data for the sources in each obscuration class
and searching for the direct signature of Fe~K$\alpha$ emission in
these composite X-ray spectra.

%
%
\begin{figure}
\centerline{\includegraphics[angle=0,width=9.0cm]{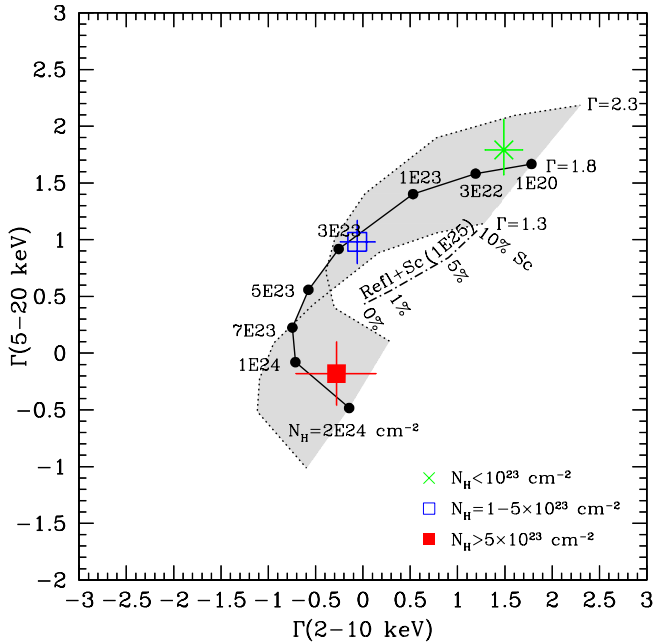}}
\vspace{-0.05in} 
\figcaption{Rest-frame 5--20~keV spectral slope versus rest-frame
  2--10~keV spectral slope, showing the results of the joint X-ray
  spectral fitting of the sources in the three obscuration classes
  from Figure~5. The model tracks are the same as those shown in
  Figure~4. The average properties of the sources in the three
  obscuration classes validate the X-ray spectral analyses for
  individual sources.}  
\vspace{-0.15in}
\end{figure}

When combining different X-ray spectra it is important to take into
account the effective area and effective exposure of each source. We
used {\sc xspec} to output the unbinned X-ray spectrum of each source,
folded by its effective area and effective exposure. We then redshift
corrected each X-ray spectrum to create a rest-frame spectrum for each
source. These rest-frame spectra were combined and then binned to
increase the signal-to-noise ratio. Since the spectral response of
\chandra\ ACIS-I varies with energy, in principal it is necessary to
take into account the spectral response for each source when their
spectra are redshifted. However, in the case of our sources the
limiting factor is photon statistics rather than spectral resolution
(e.g.,\ see \S3.3 of Alexander \etal 2003b). 

Composite rest-frame 2--20~keV spectra for the three different
obscuration classes are shown in Figure~7. The signal-to-noise ratio
of these X-ray spectra is significantly higher than the individual
X-ray spectra and discrete features and the accurate shape of the
X-ray continua can be seen. The obscured AGNs ($N_{\rm
  H}>10^{23}$~cm$^{-2}$) show a clear deficit of emission at
$\simlt$~5~keV due to absorption. As expected, there is a good
correspondence between the amount of absorption and obscuration class.
We plotted the model spectra determined from the joint X-ray spectral
fitting (see Figure~6 and \S3.3) and normalised them to the composite
X-ray spectra by eye; see Figure~7. Reassuringly, the model tracks
provide a good description of the composite X-ray spectra, validating
the results from the X-ray spectral analyses (Figures~3--6) and
showing that the variations in the properties of the individual X-ray
spectra are due to absorption.

An emission feature at the rest-frame energy of Fe~K$\alpha$ is seen
in the composite spectrum of the most heavily obscured AGNs ($N_{\rm
  H}>5\times10^{23}$~cm$^{-2}$); see Figure~7. The rest-frame
equivalent width of this feature ($\approx$~1~keV) is consistent with
that expected for Compton-thick or near Compton-thick AGNs (e.g.,\
Bassani \etal 1999) and suggests that some of the X-ray emission is
reflected and/or scattered (e.g.,\ Matt, Brandt, \& Fabian 1996). The
rest-frame energy of this feature is closer to 6.7~keV than 6.4~keV,
possibly indicating that it is produced in a warm scattering medium
rather than from cold reflection, or that contributions from both are
present. However, given the poor signal-to-noise ratio and effective
spectral resolution of the data, we cannot distinguish between these
possibilities (e.g.,\ compare the different signal-to-noise ratio
spectra of NGC~1068 and NGC~2992 in Figure~4 of Turner \etal 1997a).
If all of the X-ray emission was reflected/scattered then we would not
be able to get an accurate estimate of the intrinsic X-ray luminosity
of the AGNs in this obscuration class. However, at least for the
properties of our model, a significant fraction of the direct X-ray
emission is seen at $\simgt10$~keV, providing a reasonable constraint
on the underlying AGN luminosity. There is ambiguous evidence for
Fe~K$\alpha$ in the composite spectra of the $N_{\rm
  H}<5\times10^{23}$~cm$^{-2}$ sources, potentially indicating the
presence of Compton-thick or near Compton-thick AGN emission that is
additionally obscured by starburst regions (e.g.,\ Fabian \etal 1998).
Improved photon statistics are required to test this scenario; see
\S4.4.

In Figure~7 there appears to be an excess of $<4$~keV emission with
respect to the model in the composite spectrum of the most heavily
obscured sources ($N_{\rm H}>5\times10^{23}$~cm$^{-2}$), which may be
due to star formation. The extrapolated $\approx$~0.5--8.0~keV
luminosity ($\approx$~10$^{42}$~erg~s$^{-1}$) is consistent with the
X-ray properties of the SMGs individually classified as starburst
galaxies; see Table~1 and \S3.2. We further investigate the X-ray
emission from star formation in \S4.2.

\subsection{Absorption Correction}

Since accurate absorption corrections are challenging to determine
even for well-studied obscured AGNs in the local Universe (e.g.,\
Turner \etal 1997b; Bassani \etal 1999; Matt \etal 2000), the
absorption corrections for our X-ray faint SMGs will be somewhat
uncertain.  Thankfully, the composite X-ray spectra have shown that
our AGN model gives a good characterisation of the X-ray properties of
the AGN-classified SMGs and therefore provides a good base for
determining plausible absorption corrections. In Table~2 we show our
estimated unabsorbed X-ray luminosities for each of the AGN-classified
SMGs using our AGN model and the column-density constraints from
Figures 4--5. Although the absorption corrections for the most heavily
obscured objects ($N_{\rm H}>5\times10^{23}$~cm$^{-2}$) are the most
uncertain, they are consistent with those previously estimated for
similar sources with comparable column densities and
reflection/scattering components (e.g.,\ Iwasawa \etal 2001, 2005b;
Fabian \etal 2003; Wilman \etal 2003).  Perhaps reassuringly, the mean
unabsorbed luminosities of the $N_{\rm H}>5\times10^{23}$~cm$^{-2}$
sources are consistent with those found for the $N_{\rm
  H}=$~1--5~$\times10^{23}$~cm$^{-2}$ sources ($L_{\rm
  X}\approx6\times10^{43}$~erg~s$^{-1}$ versus $L_{\rm
  X}\approx5\times10^{43}$~erg~s$^{-1}$), as expected if the differing
amounts of absorption are primarily due to the orientation of the
obscuration with respect to the central source (i.e.,\ the unified AGN
model; Antonucci 1993). To provide a comparison to further studies in
the literature (e.g.,\ Bautz \etal 2000; Mainieri \etal 2002; Szokoly
\etal 2004; Severgnini \etal 2005), we also calculated unabsorbed
X-ray luminosities under the assumption of a simple absorbed power-law
model (i.e.,\ without the reflection and scattering components); see
Table~2.  However, we note that this model is unable to explain the
presence of Fe~K$\alpha$ emission in the most heavily obscured AGNs.

%
%
\begin{figure*}
  \centerline{\includegraphics[angle=0,width=15cm]{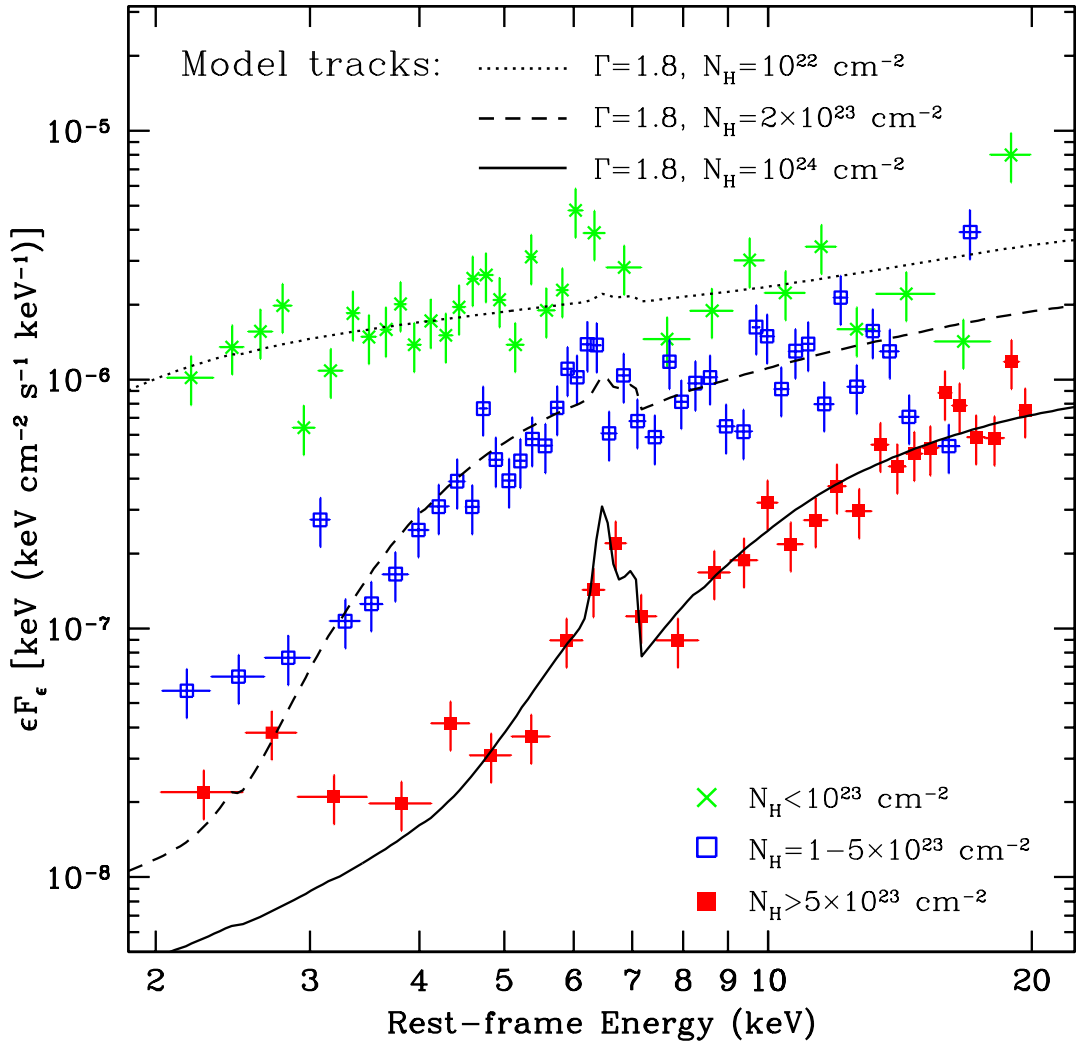}}
  \vspace{0.02in} \figcaption{Composite rest-frame 2--20~keV spectra
    for each obscuration class, as indicated; see \S3.3. The spectra
    are normalised to the average flux density of each obscuration
    class (with the exception of the the $N_{\rm H}<10^{23}$~cm$^{-2}$
    sources, which have been scaled by a factor of 2 for presentation
    purposes) and are binned at 20~counts per bin. The total number of
    counts for each obscuration class are $\approx$~690 ($N_{\rm
      H}<10^{23}$~cm$^{-2}$), $\approx$~990 ($N_{\rm
      H}=$~1--5~$\times10^{23}$~cm$^{-2}$), and $\approx$~580 ($N_{\rm
      H}>5\times10^{23}$~cm$^{-2}$). The different line styles show
    the X-ray spectra for our adopted model for different amounts of
    X-ray absorption; the absorbing column density is taken from the
    joint X-ray spectral fitting for each obscuration class (see
    Figure~6). The Fe~K$\alpha$ line is at 6.4~keV in the model
    spectra; the apparent emission feature bluewards of the
    Fe~K$\alpha$ line is due to the supposition of the different
    continuum components and the Fe absorption edge at
    $\approx$~7.1~keV. The apparent excess of $<4$~keV emission, with
    respect to the model, in the composite spectrum of the most heavily
    obscured sources ($N_{\rm H}>5\times10^{23}$~cm$^{-2}$) may be due
    to star formation (see \S3.4).} 
\end{figure*}

Although our X-ray observations probe rest-frame energies
$\simgt$~20~keV, it is possible that a large fraction of the X-ray
emission is still hidden (e.g.,\ much of the observed emission could
be scattered rather than direct). More sensitive X-ray observations,
and observations at $>$~10--20~keV (observed-frame energies), would be
able to test this scenario (e.g.,\ see Franceschini \etal 2000;
Iwasawa \etal 2001 for the brighter but similarly luminous galaxy
IRAS~09104+4109). Significant high-energy constraints could become
available with NuSTAR as soon as 2009, before further constraints with
\conx\ and \xeus\ in $>$~10~yrs time (see \S4.4).\footnote{On the
  basis of the expected $\approx$~3~$\sigma$ sensitivity limits for an
  ultra-deep $\approx$~1~Ms exposure (10--40~keV fluxes of
  $\approx10^{-14}$~erg~cm$^{-2}$~s$^{-1}$; D.~Stern, private
  communication), NuSTAR should detect emission down to $L_{\rm
    X}\approx10^{44}$--$10^{45}$~erg~s$^{-1}$ at $z\approx$~1--3; see
  http://www.nustar.caltech.edu/ for further details.} An alternative
approach is to seek independent estimates of the AGN luminosity at
other wavelengths (e.g.,\ via the [OIII]$\lambda$5007 emission line or
any strong near-IR to mid-IR high-excitation lines; Mulchaey \etal
1994; Bassani \etal 1999; Sturm \etal 2002). Only one of our
AGN-classified SMGs has an independent AGN luminosity constraint
([OIII] emission is detected from CXOHDFN~J123549.4+621536; T.~Takata,
in preparation).  The predicted rest-frame 0.5--8.0~keV luminosity of
this source ($L_{\rm X}\approx$~0.1--10~$\times10^{44}$~erg~s$^{-1}$,
determined using the [OIII]--X-ray relationship given in Bassani \etal
1999) is consistent with our estimate ($L_{\rm
  X}\approx10^{44}$~erg~s$^{-1}$); however, there are large
uncertainties and this estimate is likely to increase when the effect
of host-galaxy absorption towards the [OIII] emission-line region is
taken into account. Further independent estimates of the AGN
luminosity in SMGs would provide tighter constraints on the amount of
nuclear absorption; see also \S4.3.

\newpage

%
\section{Discussion}
%

The analyses in \S3 have provided a comprehensive characterisation of
the X-ray properties of radio-selected spectroscopically identified
SMGs with $f_{\rm 850\mu m}\simgt$~4~mJy. These properties can be used
to explore the role of AGN activity in these galaxies. In \S4.1 we
further investigate the AGN properties of our SMG sample, in \S4.2 we
estimate the contribution that AGN activity makes to the bolometric
luminosity, in \S4.3 we consider the constraints on black-hole growth,
and in \S4.4 we discuss the prospects for deeper X-ray observations
and explore the scientific potential of the next generation of X-ray
observatories for the study of SMGs.

\subsection{AGN Activity in the SCUBA Galaxy Sample}

The X-ray properties of the AGN-classified SMGs in our sample are
generally consistent with those of nearby luminous AGNs (i.e.,\
$\Gamma\approx1.8\pm0.5$, $N_{\rm
  H}\approx$~$10^{20}$--$10^{24}$~cm$^{-2}$, and $L_{\rm
  X}\approx$~$10^{43}$--$10^{44.5}$~erg~s$^{-1}$; e.g.,\ Smith \& Done
1996; Turner \etal 1997b; Maiolino \etal 1998; Risaliti \etal 1999);
see \S3 and Table~2. The majority ($\approx$~80\%) of the AGNs are
heavily obscured ($N_{\rm H}\simgt$~$10^{23}$~cm$^{-2}$), four of
which have $L_{\rm X}\simgt10^{44}$~erg~s$^{-1}$ and could be
considered obscured (Type 2) quasars; see Bautz \etal (2000) and
Mainieri \etal (2005a) for other examples of obscured quasars that are
bright at submm wavelengths.  With the exception of completely
Compton-thick AGNs (i.e.,\ $N_{\rm H}\approx10^{25}$~cm$^{-2}$), the
overall column density distribution would be roughly similar to that
found for nearby AGNs (e.g.,\ Risaliti \etal 1999; Salvati \& Maiolino
2000). The lack of obvious X-ray emission from AGN activity in the
X-ray classified starbursts could be because the AGNs are of a low
luminosity ($L_{\rm 0.5-8.0~keV}\simlt10^{42}$~erg~s$^{-1}$) or are
completely Compton thick ($N_{\rm
  H}\approx$~$10^{25}$~cm$^{-2}$).\footnote{Of course some of the X-ray
  classified starbursts may not host AGN activity.  However, we note
  that the similar $z=2.285$ galaxy FSC~10214+4724 hosts an AGN that
  is weak at X-ray energies and would have been classified as a
  starburst galaxy in our study (Alexander \etal 2005b).}
Interestingly, if the latter is assumed then the estimated column
density distribution would be remarkably similar to that found for
nearby AGNs.

The obscured AGN fraction in our sample ($\approx$~80\% have $N_{\rm
  H}\simgt$~$10^{23}$~cm$^{-2}$) is larger than that found from
identification studies of the $\approx$~2--8~keV background population
(e.g.,\ Ueda \etal 2003 find similar numbers of obscured and
unobscured AGNs; see also Barger \etal 2002; Mainieri \etal 2002;
Szokoly \etal 2004). Although this may suggest a prevalence for
obscured activity in the SMG population, selection effects are also
likely to be at least partially responsible. For example, due to the
comparatively high redshifts of our sources we probe considerably
higher rest-frame energies than those for the bulk of the
spectroscopically identified sources in deep X-ray surveys [e.g.,\
$\approx$~1.5--24~keV (at $z\approx$~2) as compared to
$\approx$~0.9--14~keV (at $z\approx$~0.7) at observed energies of
0.5--8.0~keV], allowing higher column densities to be penetrated.
Further $z>1$ obscured AGNs are likely to have been detected in X-ray
surveys but lack spectroscopic redshifts due to the faintness of their
optical counterparts (e.g.,\ Alexander \etal 2001; Barger \etal 2002;
Fiore \etal 2003; Treister \etal 2004; Mainieri \etal 2005b). These
optically faint X-ray sources are the dominant population of obscured
AGNs at $z>1$ and may account for as much as $\approx$~50\% of the
AGNs detected in deep X-ray surveys (Alexander \etal 2002). Complete
spectroscopic identification of these sources is required to determine
the obscured to unobscured AGN ratio at $z>1$ and reveal whether our
X-ray detected SMGs are preferentially more obscured than the general
$z>1$ X-ray source population.

\subsection{What Powers SCUBA Galaxies?} 

AGN activity clearly plays an important role in the spectroscopically
identified SMG population but does it dominate the bolometric output?
The answer to this key question can have important implications for
the formation and evolution of massive galaxies and the growth of
massive black holes (see \S1). As found for most dusty luminous
galaxies, the luminosity of SMGs is likely to peak at far-IR
wavelengths and have components of both AGN and star-formation
activity (e.g.,\ Sanders \& Mirabel 1996; Genzel \& Cesarsky 2000).
Many studies have suggested that star formation accounts for a large
fraction of the bolometric output of SMGs (e.g.,\ Frayer \etal 1998,
2004; Ivison \etal 2002, 2004; Alexander \etal 2003b; Almaini \etal
2003; Chapman \etal 2004a; Egami \etal 2004; Swinbank \etal 2004) but
none has resolved the rest-frame far-IR emission. The most direct
study to date is that of Chapman \etal (2004a) who showed that
$\approx$~70\% of radio-identified SMGs have radio emission extended
on $\approx$~10~kpc scales, implying that the rest-frame far-IR
emission is also extended over these scales. However, these results do
not rule out the presence of energetically significant AGN components
on smaller scales. The important contribution our study can make to
this debate is in providing direct estimates of the bolometric
contribution from AGN activity.

In Figure~8 we show the rest-frame far-IR luminosity versus unabsorbed
0.5--8.0~keV luminosity for the SMGs and compare them to well-studied
starburst galaxies and AGNs drawn from the literature. The shaded
region indicates the typical range of luminosity ratios for the
well-studied quasars of Elvis \etal (1994) and provides an indication
of the location of AGN-dominated sources on this figure. None of the
far-IR luminous SMGs ($L_{\rm FIR}\simgt10^{12}$~\Lsolar) lies within
the dark-shaded region, indicating that they are comparatively weak at
X-ray energies. The difference between the median X-ray-to-far-IR
luminosity ratios of the AGN-classified SMGs (${L_{\rm X}}\over{L_{\rm
    FIR}}$$\approx$~0.004) and the quasars (${L_{\rm X}}\over{L_{\rm
    FIR}}$$\approx$~0.05) suggest that the AGN activity in the SMGs
contributes, on average, only $\approx$~8\% of the far-IR emission.
This comparison indicates that star formation typically dominates the
bolometric output of these SMGs. However, if the SMGs have an AGN
dust-covering factor $\approx$~12 times larger than found in the
quasars then they could potentially be AGN dominated (i.e.,\ a larger
amount of dust could be heated for a given central source luminosity).
Although the large fraction of obscured AGNs in our sample suggests
that the dust-covering factor is large in these SMGs, the
dust-covering factor of quasars is uncertain and may depend on the
evolutionary state or nature of individual objects (e.g.,\ Haas \etal
2003; Page \etal 2004).

We can take a different approach and compare our SMGs to the
literature galaxies, many of which host an obscured AGN and may be
physically similar to the SMGs; see Figure~8. The complication of this
comparison is that the relative contribution from AGN and
star-formation activity in these galaxies is often poorly constrained.
Even so, we can gain some insight into the relative dominance of AGN
and star-formation activity by ``calibrating'' Figure~8 using the
\iso\ mid-IR spectral diagnostics of Rigopoulou \etal (1999) and Tran
\etal (2001); see also Genzel \etal (1999) and Lutz, Veilleux, \&
Genzel (1999). On the basis of these studies, we have found that the
%
%
\begin{figure*}[t]
  \centerline{\includegraphics[angle=0,width=15cm]{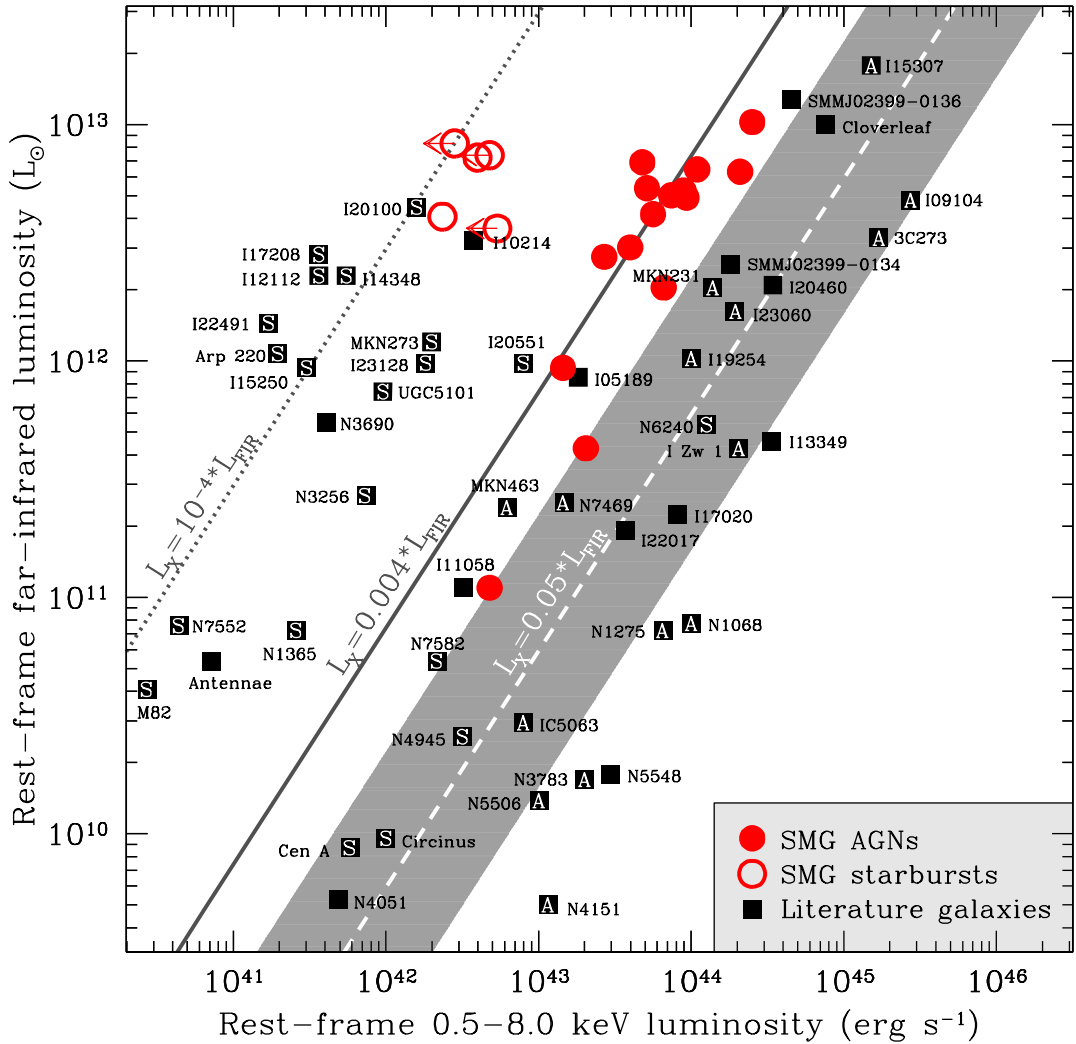}}
  \vspace{0.02in} \figcaption{Rest-frame far-IR versus unabsorbed
    0.5--8.0~keV luminosity for our SMGs (circles) and galaxies
    selected from the literature (squares; individual galaxies are
    labelled to allow objects to be compared to the SMGs). Sources are
    further highlighted as follows: AGN-classified SMGs (filled
    circles), starburst-classified SMGs (open circles), literature
    galaxies (filled squares); literature galaxies classified as
    AGN-dominated or star-formation dominated by Rigopoulou \etal
    (1999) and Tran \etal (2001) are additionally indicated by an
    ``A'' or ``S'', respectively. The diagonal lines show ratios of
    constant X-ray-to-far-IR luminosity: the dotted line shows the
    mean luminosity ratio of the starburst-classified SMGs, the solid
    line shows the median luminosity ratio of the AGN-classified SMGs,
    and the dashed line shows the median luminosity ratio for the
    quasars studied by Elvis \etal (1994).  The shaded region
    indicates the standard-deviation in the luminosity ratio of the
    quasars studied by Elvis \etal (1994). The X-ray and infrared data
    for the literature galaxies were taken from Sanders \& Mirabel
    (1996), Hughes \etal (1997), Bassani \etal (1999), Bautz \etal
    (2000), Matt \etal (2000), Risaliti \etal (2000), Iwasawa (2001),
    Iwasawa, Fabian, \& Ettori (2001), Gallagher \etal (2002), Lira
    \etal (2002), Verma \etal (2002), Braito \etal (2003, 2004),
    Franceschini \etal (2003), Sanders \etal (2003), Alexander \etal
    (2005b), Brandt \& Hasinger (2005), Iwasawa \etal (2005b), and the
    NASA Extragalactic Database (NED).  The literature data have been
    converted to the cosmology used here and, when appropriate, the
    X-ray data have been converted to the 0.5--8.0~keV band assuming
    $\Gamma=1.8$ for the AGNs and $\Gamma=2.0$ for the starburst
    galaxies. Although the unabsorbed X-ray luminosities are presented
    whenever possible, in some sources the intrinsic power of the AGN
    is unknown (e.g.,\ the possible Compton-thick AGN in
    FSC~10214+4724; Alexander \etal 2005b) and the observed X-ray
    luminosity is shown.  In the absence of far-IR data, far-IR
    luminosities were estimated from the 1.4~GHz luminosity density
    under the assumption of the radio-to-far-IR relationship.}
\end{figure*}

\noindent threshold between the AGN-dominated and star-formation
dominated literature galaxies is ${{L_{\rm X}}\over{L_{\rm
      FIR}}}\approx$~0.004; of the 35 galaxies in Rigopoulou \etal
(1999) and Tran \etal (2001), all of the 17 AGN-dominated galaxies
have ${{L_{\rm X}}\over{L_{\rm FIR}}}>$~0.004 while 13 of the 18
star-formation dominated galaxies have ${{L_{\rm X}}\over{L_{\rm
      FIR}}}<$~0.004 (see Figure~8).  This indicates that
AGN-dominated galaxies can lie outside of the shaded region in
Figure~8 and suggests that the Elvis \etal (1994) bolometric
correction is unlikely to be applicable for all sources.  The
implication of these results is that some of the SMGs could be
predominantly powered by AGN activity even though the majority are
likely to be star-formation dominated. However, since different
techniques often yield contrasting results (e.g.,\ Farrah \etal 2003),
and the exact contributions from AGN and star-formation activity are
uncertain {\it even} in extremely well-studied local galaxies (e.g.,\
Spoon \etal 2000; Lutz \etal 2003; Iwasawa \etal 2005a), this
conclusion can only be considered qualitative. Deep mid-IR
spectroscopy with the \spitzer\ telescope should provide important
additional insight into the relative contributions from AGN and
star-formation activity in these SMGs (e.g.,\ Lutz \etal 2005 for the
first constraints).  However, significantly improved constraints are
unlikely to be achieved before the advent of the Atacama Large
Millimeter Array (ALMA), which will offer spatial resolution
constraints at submm wavelengths on $\approx$~100~pc scales for
galaxies at $z\approx$~2.\footnote{See http://www.alma.nrao.edu/ for
  more details on ALMA.}

In Figure~8 we also show the X-ray and far-IR luminosities of the
starburst-classified SMGs. The mean X-ray-to-far-IR luminosity ratio
of these sources (${{L_{\rm X}}\over{L_{\rm FIR}}}\approx10^{-4}$,
when the X-ray stacking result of the three X-ray undetected SMGs is
included; see \S2.2) is consistent with that expected from star
formation if the X-ray emission is dominated by HMXBs (Persic \etal
2004; see also Figure~2b and \S3.2). Under this assumption the X-ray
derived star-formation rates ($\approx$~1300--2700~\Msolar yr$^{-1}$)
are in good agreement with those determined using other techniques
(e.g.,\ Swinbank \etal 2004; Chapman \etal 2005). The mean
X-ray-to-far-IR luminosity ratio is $\approx$~4 times lower than that
found for typical starburst galaxies in the local Universe. However,
as argued in Persic \etal (2004; see also Franceschini \etal 2003),
the X-ray emission from typical starburst galaxies will be dominated
by long-lived LMXBs, which will be comparatively weak in intense
star-forming galaxies. Of course, the weak X-ray emission from these
sources could also imply that luminous completely Compton-thick AGNs
are present (e.g.,\ Iwasawa \etal 2005a).

\subsection{Growth of Black Holes in Massive Star-Forming Galaxies}

In Alexander \etal (2005a) we used the results presented here to
constrain the growth of black holes in massive star-forming galaxies.
We showed that the AGN fraction in the bright SMG population is
$>$~38$^{+12}_{-10}$\%, when corrected for bright SMGs without
spectroscopic redshifts, and argued that this indicates that their
black holes are almost continuously growing throughout vigorous
star-formation episodes. The most likely catalyst for this activity
appeared to be galaxy major mergers (e.g.,\ Chapman \etal 2003c;
Conselice \etal 2003). Shortly after the completion of our study
sophisticated hydrodynamical simulations of galaxy major mergers,
taking into account the growth of both the black hole and stellar
components, were published (e.g.,\ Di Matteo \etal 2005; Springel
\etal 2005). The simulation results are in good agreement with the
conclusions of Alexander \etal (2005a), with the peak epoch of star
formation corresponding to a heavily obscured rapid black-hole growth
phase, which is ultimately proceeded by an unobscured quasar phase
(Hopkins \etal 2005). However, the black holes of the most massive
galaxies in these simulations (which are probably most similar to the
SMGs) are up-to an order of magnitude more massive than those
estimated from our X-ray luminosities under the assumption of
Eddington-limited accretion ($M_{\rm BH (Edd)}\simlt10^8$~\Msolar).
Our results can be better reconciled with the simulations if either
the black holes in the AGN-classified SMGs are accreting at
sub-Eddington rates, a large fraction of the AGN activity is hidden at
rest-frame energies of $\approx$~20~keV (see \S3.5), or the
X-ray-to-bolometric luminosity ratio is significantly lower than that
typically assumed for AGNs (i.e.,\ Elvis \etal 1994; see Table~1 of
Alexander \etal 2005a for further details). However, there is further
observational evidence that the masses of the black holes in SMGs may
be comparatively modest and in good agreement with our estimate.
Rest-frame optical spectra of SMGs show that when broad emission lines
are detected they are often comparatively narrow (typical full width
at half-maximum velocities of $\approx$~1000--3000~km~$^{-1}$; Vernet
\& Cimatti 2001; Smail \etal 2003; Swinbank \etal 2004, 2005;
T.~Takata \etal, in preparation).  Under the assumption that the
dynamics of the broad-line regions in these sources are dominated by
the gravity of the central black hole, these comparatively narrow
emission line widths suggest typical black-hole masses of
$\simlt10^{8}$~\Msolar\ for mass accretion rates of
$\simlt$~1~$M_{\odot}$~yr$^{-1}$ (Alexander \etal 2005a; e.g.,\ McLure
\& Dunlop 2004). Although it is premature to draw conclusions before
more accurate black-hole mass constraints are secured, we note that
other models have suggested that the masses of the black holes in SMGs
are comparatively modest (e.g.,\ Archibald \etal 2002; Kawakatu \etal
2003; Granato \etal 2004a,b). In particular, the physically motivated
models of Granato \etal bear a striking similarity to our results (see
Figures~2 \& 3 in Granato \etal 2004b for the $M_{\rm
  vir}=10^{12.4}$~\Msolar\ model).

\subsection{Prospects for Deeper X-ray Observations}

The quality of the individual X-ray spectra are restricted by poor
photon statistics. However, the composite X-ray spectra in \S3.4
demonstrate what could be achieved for individual sources with
extremely long \chandra\ exposures (up to $\approx$~12~Ms). With
individual X-ray spectra of this quality we would be able to
significantly improve the constraints for individual sources (e.g.,\
$\Gamma$, $N_{\rm H}$, the strength of Fe~K$\alpha$ emission,
intrinsic X-ray luminosity, measure soft X-ray excesses due to
star-formation activity). The increased photon statistics would also
allow for basic X-ray spectral analyses (up to $\approx$~200 counts)
of the starburst-classified SMGs.

The major scientific thrust of the next generation of X-ray
observatories (\conx\ and \xeus) is high-resolution spectroscopy of
faint X-ray sources.\footnote{See
  http://constellation.gsfc.nasa.gov/science/index.html and
  http://www.rssd.esa.int/XEUS/ for more details.} In order to
investigate the scientific potential of these observatories for the
study of SMGs, we have simulated a 300~ks \xeus\ exposure of one of
the most heavily obscured sources (SMMJ~123622.6+621629 at $z=2.466$)
using {\sc xspec}; see Figure~9. An Fe~K$\alpha$ emission line is
easily identifiable at rest-frame 6.4~keV, indicating that this is a
Compton-thick or near Compton-thick AGN; similar results can be
obtained with \conx\ for sources $\approx$~10 times brighter. The
quality (signal-to-noise ratio and spectral resolution) of the
simulated X-ray spectrum exceeds that of our composite X-ray spectra
(compare with Figure~7). Both \conx\ and \xeus\ will also have the
capability to detect sources at $>$~10--20~keV (observed-frame
energies) and place crucial AGN luminosity constraints at very high
X-ray energies (see Footnote 6 for NuSTAR constraints); see \S3.5.

The increased photon statistics and spectral resolution of the
next-generation X-ray observatories should allow for the detection of
weaker AGN features. For example, some local AGNs and powerful quasars
have shown evidence for large-scale outflowing material, probably due
to accretion-disk winds (e.g.,\ Chartas \etal 2002; Kaspi \etal 2002;
Ogle \etal 2003).  These outflows provide an efficient method for
distributing high-metallicity gas into the intergalactic medium and
may be responsible for producing the M--${\sigma}$ relationship seen
in local galaxies (e.g.,\ Silk \& Rees 1998; Fabian 1999; King 2003;
Di Matteo \etal 2005). If SMGs are the precursors to luminous quasar
activity, and the ancestors of nearby
%
%
\begin{figure*}
\centerline{\includegraphics[angle=-90,width=15.0cm]{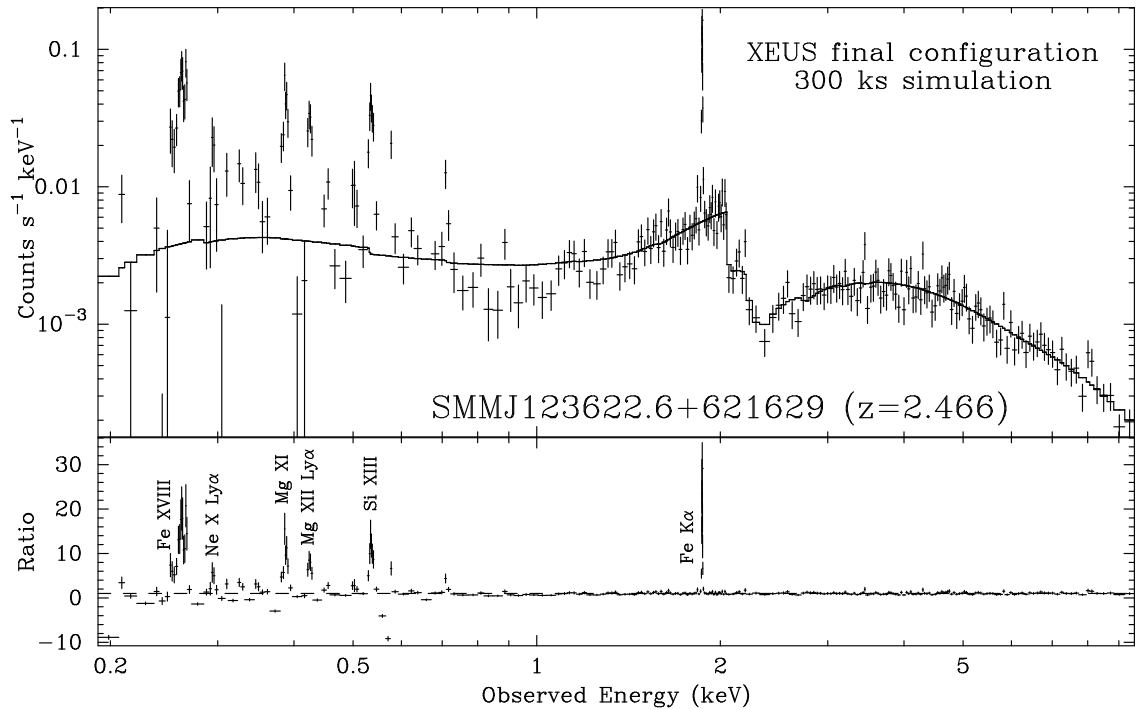}}
\vspace{0.02in}
\figcaption{Simulated 300~ks final-configuration \xeus\ spectrum of
  the heavily obscured source SMMJ~123622.6+621629 (top panel). The
  spectrum was simulated using our adopted AGN model (see \S2.3) with
  $\Gamma=1.8$ and $N_{\rm H}=1.5\times10^{24}$~cm$^{-2}$, and
  additionally includes an accretion-disk wind outflow [taken from NGC~1068 (Ogle
  \etal 2003) and scaled to the Fe~K$\alpha$ flux of our source]; see
  \S4.4. The solid line indicates the input model with the
  emission-line components removed to highlight the emission-line
  features. The bottom panel shows the ratio of the simulated data to
  the model.  Rest-frame 6.4~keV Fe~K$\alpha$ is easily identified
  (at $\approx$~1.8~keV) and indicates that this source is a
  Compton-thick or near Compton-thick AGN. The $<$~1~keV emission
  features are due to the outflowing accretion-disk wind. Six of the
  strongest emission lines are labelled in the bottom panel (see Ogle
  \etal 2003 for further information and line identifications).} 
\end{figure*}

\noindent massive galaxies, then they are likely to produce
significant emission-line outflows (see Smail \etal 2003; Chapman
\etal 2004b; Swinbank \etal 2005 for evidence at optical/near-IR
wavelengths). We included the emission-line outflow parameters found
for NGC~1068 (Ogle \etal 2003) to our simulated 300~ks \xeus\
spectrum, scaled to the Fe~K$\alpha$ flux of our source; see Figure~9.
The emission-line signatures of the outflow are identifiable in the
simulated spectrum at $\simlt1$~keV, allowing basic constraints to be
placed on the properties of the outflowing material (e.g.,\ the mass
outflow rate).  These constraints could prove to be crucial in our
understanding of the connection between AGN and star-formation
activity at high redshift, and help to explain the properties of
nearby massive galaxies.

%
\section{Conclusions}
%

Using the powerful combination of ultra-deep X-ray observations and
deep optical spectroscopic data we have placed constraints on the
X-ray properties of 20 radio-selected spectroscopically identified
SMGs with $f_{\rm 850\mu m}\simgt$~4~mJy. Our key results are the
following:

\begin{enumerate}

\item Seventeen of the 20 SMGs in our sample are detected at X-ray
  energies. From a classification of their X-ray properties, we found
  the X-ray emission to be dominated by AGN activity in 15 sources
  ($\approx$~75\%); the X-ray emission from the other 5 sources is
  likely to be dominated by star-formation activity. See \S3.2.

\item Using a variety of X-ray spectral analyses we found that the
  properties of the AGNs are generally consistent with those of nearby
  luminous AGNs (i.e.,\ $\Gamma\approx1.8\pm0.5$, $N_{\rm
    H}\approx$~$10^{20}$--$10^{24}$~cm$^{-2}$, and $L_{\rm
    X}\approx$~$10^{43}$--$10^{44.5}$~erg~s$^{-1}$); see \S3 and
  Table~2. The majority ($\approx$~80\%) of the AGNs are heavily
  obscured ($N_{\rm H}\simgt$~$10^{23}$~cm$^{-2}$), four of which have
  $L_{\rm X}\simgt10^{44}$~erg~s$^{-1}$ and could be considered
  obscured quasars. The estimated column-density distribution is
  roughly similar to that found for nearby AGNs. See \S3.3 and \S4.1.

\item We constructed rest-frame 2--20~keV composite spectra for three
  different obscuration classes ($N_{\rm H}<10^{23}$~cm$^{-2}$,
  $N_{\rm H}=$~1--5~$\times10^{23}$~cm$^{-2}$, and $N_{\rm
    H}>5\times10^{23}$~cm$^{-2}$). These composite spectra revealed
  features not observed in the individual X-ray spectra and an
  $\approx$~1~keV equivalent width Fe~K$\alpha$ emission line is seen
  in the composite X-ray spectrum of the most heavily obscured AGNs,
  suggesting Compton-thick or near Compton-thick absorption. The good
  agreement between our model tracks and the composite X-ray spectra
  validate the X-ray spectral analyses of individual sources. See
  \S3.4. 

\item Taking into account the effects of absorption, we found that the
  average X-ray to far-IR luminosity ratio of the AGN-classified
  sources (${L_{\rm X}}\over{L_{\rm FIR}}$=~0.004) is approximately
  one order of magnitude below that found for typical quasars. This
  result suggests that intense star-formation activity (of order
  $\approx$~1000~\Msolar yr$^{-1}$) dominates the bolometric output of
  spectroscopically identified SMGs. Possible biases in the selection
  of this sample (see \S2.1) would suggest that, if anything, they
  should have a higher AGN contribution than spectroscopically
  unidentified SMGs. We also investigated the possibility that the
  X-ray to far-IR luminosity ratio for the AGNs in SMGs is
  intrinsically less than that found for typical quasars and
  postulated that some SMGs may be AGN dominated. See \S4.2.

\item We found that the X-ray emission from the starburst-classified
  SMGs is consistent with that expected from HMXBs. The X-ray derived
  star-formation rates ($\approx$~1300--2700~\Msolar yr$^{-1}$) are in
  good agreement with those determined using other techniques.  See
  \S4.2.

\item We find good agreement between our overall picture for the
  growth of black holes in massive galaxies and the results from
  recent hydrodynamic simulations. We provide support that the black
  holes in our AGN-classified SMGs are $\simlt10^{8}$~\Msolar. See
  \S4.3.

\item We demonstrated that the next generation of X-ray observatories
  (\conx\ and \xeus) will place significantly improved X-ray spectral
  constraints on individual sources. In addition to constraining the
  absorption properties of the AGN components we found that these
  X-ray observatories have the potential to detect the presence and
  properties of outflowing material. These observations could prove to
  be crucial in our understanding of the evolution of massive
  galaxies. See \S4.4.

\end{enumerate}


%
\section*{Acknowledgments}
%

We acknowledge support provided by the Royal Society (DMA, IS), PPARC
(FEB), NASA \#9174 (SSC), NSF award AST-0205937, the Research
Corporation, and the Alfred P. Sloan Foundation (AWB), NSF CAREER
award AST-99833783, and CXC grant G02-3187A (WNB). We are grateful to
Stefan Immler for performing the X-ray stacking analyses on the X-ray
undetected \scuba\ galaxies. We thank Omar Almaini, Tiziana Di Matteo,
Andy Fabian, Reinhard Genzel, John Grimes, Gunther Hasinger, Kazushi
Iwasawa, Pat Ogle, Maurizio Paolillo, Andy Ptak, Nick Scoville, Dan
Stern, Tadafumi Takata, and Cristian Vignali for useful comments and
suggestions. We thank the referee for presentation suggestions. This
research has made use of the NASA/IPAC Extragalactic Database (NED)
which is operated by the Jet Propulsion Laboratory, California
Institute of Technology, under contract with the National Aeronautics
and Space Administration.

%

\newpage

%
%

\newpage

%
%

\clearpage
\LongTables 
\begin{landscape}
\tabletypesize{\small}
\begin{deluxetable}{cccccccccccccc}
\tablecaption{Basic properties of the \scuba\ galaxies}
\tablehead{
\multicolumn{4}{c}{SCUBA Galaxy Properties} &
\colhead{Radio} &
\colhead{Far-IR}  &
\colhead{X-ray}  &
\multicolumn{2}{c}{X-ray} &
\colhead{Band}                  &
\colhead{Effective}             &
\multicolumn{3}{c}{X-ray Source Properties}  \\
\colhead{$\alpha_{2000}$$^{\rm a}$}       &
\colhead{$\delta_{2000}$$^{\rm a}$}       &
\colhead{$S_{\rm 850\mu m}$$^{\rm a}$}&
\colhead{$z^{\rm a}$}        &
\colhead{$L_{\rm 1.4~GHz}$$^{\rm b}$}              &
\colhead{$L_{\rm FIR}$$^{\rm c}$}              &
\colhead{ID$^{\rm d}$}               &
\colhead{$\alpha_{2000}$$^{\rm d}$}       &
\colhead{$\delta_{2000}$$^{\rm d}$}       &
\colhead{Ratio$^{\rm e}$}       &
\colhead{$\Gamma$$^{\rm e}$}     &
\colhead{$f_{\rm 0.5-8 keV}$ $^{\rm f}$} &
\colhead{$L_{\rm 0.5-8 keV}$$^{\rm g}$} &
\colhead{Class$^{\rm h}$}}
\startdata
12 35 49.44& $+$62 15 36.8& 8.3$\pm$2.5& 2.203& 24.38 & 12.72 & 35 & 12 35 49.44 & $+$62 15 36.9 &    1.64$^{+0.67}_{-0.50}$ & 0.38$^{+0.33}_{-0.31}$&   1.34 &   43.6 & AGN \\
12 35 53.26& $+$62 13 37.7& 8.8$\pm$2.1& 2.098& 24.22 & 12.56 &\dots & \dots          & \dots  &    \dots                  & \dots                 &$<$0.19 &$<$42.7 & S/Burst? \\
12 35 55.14& $+$62 09 01.7& 5.4$\pm$1.9& 1.875& 24.67 & 13.01 & 44 & 12 35 55.13 & $+$62 09 01.7 &    1.53$^{+0.27}_{-0.24}$ & 0.44$^{+0.15}_{-0.15}$&  11.25 &   44.4 & AGN \\
12 36 00.15& $+$62 10 47.2& 7.9$\pm$2.4& 2.002& 24.52 & 12.87 & \dots & \dots          & \dots &    \dots                  & \dots                 &$<$0.19 &$<$42.7 & S/Burst? \\
12 36 06.72& $+$62 15 50.7& 4.4$\pm$1.4& 2.416& 23.97 & 12.31 & 77 & 12 36 06.70 & $+$62 15 50.7 &    0.25$^{+0.05}_{-0.04}$ & 2.17$^{+0.32}_{-0.29}$&   1.69 &   43.8 & AGN \\
\\
12 36 06.85& $+$62 10 21.4&11.6$\pm$3.5& 2.509& 24.50 & 12.84 & 79 & 12 36 06.84 & $+$62 10 21.4 &    $>$2.18                & $<$0.12               &   0.74 &   43.5 & AGN \\
12 36 16.15& $+$62 15 13.7& 5.8$\pm$1.1& 2.578& 24.39 & 12.73 & 109& 12 36 16.11 & $+$62 15 13.7 &    0.85$^{+0.20}_{-0.18}$ & 0.97$^{+0.21}_{-0.19}$&   1.02 &   43.7 & AGN \\
12 36 18.33& $+$62 15 50.5& 7.3$\pm$1.1& 1.865& 24.52 & 12.86 &SUPP& 12 36 18.40 & $+$62 15 51.2 &    \dots                  & 1.40                  &   0.18 &   42.6 & S/Burst? \\
12 36 21.27& $+$62 17 08.4& 7.8$\pm$1.9& 1.998& 24.58 & 12.92 & \dots & \dots          & \dots &    \dots                  & \dots               &$<$0.11 &$<$42.5 & S/Burst? \\
12 36 22.65& $+$62 16 29.7& 7.7$\pm$1.3& 2.466& 24.46 & 12.81 & 135& 12 36 22.66 & $+$62 16 29.8 &    $>$4.39                & $<$--0.50             &   1.02 &   43.6 & AGN \\
\\
12 36 29.13& $+$62 10 45.8& 5.0$\pm$1.3& 1.013& 23.63 & 11.97 & 158& 12 36 29.11 & $+$62 10 45.9 &    2.05$^{+0.38}_{-0.33}$ & 0.18$^{+0.16}_{-0.15}$&   2.31 &   43.1 & AGN \\
12 36 32.61& $+$62 08 00.1& 4.9$\pm$1.2& 1.993& 24.36 & 12.70 & 171& 12 36 32.59 & $+$62 07 59.8 &    2.37$^{+0.68}_{-0.53}$ & 0.05$^{+0.23}_{-0.22}$&   1.83 &   43.7 & AGN \\
12 36 34.51& $+$62 12 41.0& 5.1$\pm$1.6& 1.219& 24.27 & 12.61 & 182& 12 36 34.50 & $+$62 12 41.2 &    $<$0.41                & $>$1.63               &   0.29 &   42.4 & S/Burst? \\
12 36 35.59& $+$62 14 24.1& 4.7$\pm$1.3& 2.005& 24.35 & 12.69 & 190& 12 36 35.58 & $+$62 14 24.1 &    1.90$^{+0.30}_{-0.27}$ & 0.25$^{+0.14}_{-0.13}$&   2.52 &   43.8 & AGN (V)\\
12 36 36.75& $+$62 11 56.1& 7.0$\pm$2.1& 0.555& 22.70 & 11.04 & 194& 12 36 36.75 & $+$62 11 56.0 &    0.31$^{+0.03}_{-0.03}$ & 1.87$^{+0.09}_{-0.09}$&   3.77 &   42.7 & AGN (V)\\
\\
12 37 07.21& $+$62 14 08.1& 4.7$\pm$1.5& 2.484& 24.27 & 12.62 & 347& 12 37 07.20 & $+$62 14 07.9 &    1.62$^{+0.48}_{-0.39}$ & 0.39$^{+0.25}_{-0.23}$&   0.98 &   43.6 & AGN \\
12 37 11.98& $+$62 13 25.7& 4.2$\pm$1.4& 1.992& 24.13 & 12.48 & 368& 12 37 12.04 & $+$62 13 25.7 &    $>$4.11                & $<$--0.44             &   0.92 &   43.4 & AGN \\
12 37 12.05& $+$62 12 12.3& 8.0$\pm$1.8& 2.914& 24.10 & 12.44 & 369& 12 37 12.09 & $+$62 12 11.3 &    2.00$^{+1.45}_{-0.88}$ & 0.20$^{+0.52}_{-0.49}$&   0.38 &   43.4 & AGN \\
12 37 16.01& $+$62 03 23.3& 5.3$\pm$1.7& 2.037& 24.46 & 12.80 & 385& 12 37 16.04 & $+$62 03 23.7 &    0.42$^{+0.09}_{-0.08}$ & 1.61$^{+0.20}_{-0.17}$&   7.36 &   44.3 & AGN (V)\\
12 37 21.87& $+$62 10 35.3&12.0$\pm$3.9& 0.979& 23.29 & 11.63 & 405& 12 37 21.86 & $+$62 10 35.8 &    5.07$^{+2.43}_{-1.52}$&--0.63$^{+0.32}_{-0.35}$&   2.11 &   43.1 & AGN (V)\\
\enddata
\tablenotetext{a}{SCUBA galaxy properties, taken from Chapman et~al. (2005). Submm flux density is in units of mJy.} 
\tablenotetext{b}{Rest-frame 1.4~GHz luminosity density in logarithmic units of W Hz$^{-1}$. Calculated following equation 2 in Alexander et~al. (2003b) for $\alpha=0.8$ (the average spectral index for star-forming galaxies; e.g.,\ Yun, Reddy, \& Condon 2001) using the radio flux density from Chapman et~al. (2005).} 
\tablenotetext{c}{Rest-frame far-IR ($\lambda$~=40--120~$\mu$m) luminosity in logarithmic units of solar luminosities. Calculated from the rest-frame 1.4~GHz luminosity under the assumption of the radio to far-IR correlation with $q=2.35$ (e.g.,\ Helou, Soifer, \& Rowan-Robinson 1985).} 
\tablenotetext{d}{X-ray source identification number and X-ray source co-ordinates; taken from Alexander et~al. (2003a). ``SUPP'' refers to a source detected in the complete supplementary \chandra\ catalog; see Alexander et~al. (2003b), \S3.2 and \S3.4.2 of Alexander et~al. (2003a).} 
\tablenotetext{e}{Ratio of the count rates in the 2.0--8.0~keV and 0.5--2.0~keV bands, and the effective photon index in the 0.5--8.0~keV band (calculated from the band ratio); taken from Alexander et~al. (2003a). $\Gamma=1.4$ has been adopted for sources with poorly defined band ratios.} 
\tablenotetext{f}{Full-band flux uncorrected for Galactic or intrinsic absorption, in units of $10^{-15}$~erg~cm$^{-2}$~s$^{-1}$. Taken from Alexander et~al. (2003a,b). Upper limits are calculated following \S3.4.1 of Alexander et~al. (2003a).} 
\tablenotetext{g}{Rest-frame 0.5--8.0~keV luminosities in logarithmic units of erg~s$^{-1}$, calculated following equation 1 in Alexander et~al. (2003b) for $\Gamma=1.8$..} 
\tablenotetext{h}{Source classification based on X-ray properties; see \S3.2. ``AGN'' indicates the X-ray emission is dominated by AGN activity, ``(V)'' indicates evidence for X-ray variability, and ``S/Burst?'' indicates that the X-ray emission is probably dominated by luminous star-formation activity.} 
\end{deluxetable}
\clearpage
\end{landscape}

%
%

\clearpage
\LongTables 
\begin{landscape}
\tabletypesize{\small}
\begin{deluxetable}{cccccccccccc}
\tablecaption{Spectral fits for the AGN-classified SCUBA galaxies}
%
\tablehead{
\multicolumn{2}{c}{X-ray} &
\colhead{Rest-Frame}       &
\colhead{X-ray}                  &
\multicolumn{2}{c}{X-ray Spectral Fits} &
\colhead{}                  &
\multicolumn{4}{c}{X-ray Luminosities} \\
\colhead{$\alpha_{2000}$$^{\rm a}$}       &
\colhead{$\delta_{2000}$$^{\rm a}$}       &
\colhead{Energies$^{\rm b}$}          &
\colhead{Counts$^{\rm c}$}       &
\colhead{$\Gamma$(2--10~keV)$^{\rm d}$}          &
\colhead{$\Gamma$(5--20~keV)$^{\rm d}$}       &
\colhead{$N_{\rm H}$$^{\rm e}$}      &
\colhead{$L_{\rm 2-10 keV}$$^{\rm f}$}  &
\colhead{$L_{\rm 5-20 keV}$$^{\rm f}$}  &
\colhead{$L_{\rm 0.5-8 keV, corr}$$^{\rm g}$} &
\colhead{$L_{\rm 0.5-8 keV, pow}$$^{\rm h}$}}
\startdata
12 35 49.44 & $+$62 15 36.9 & 1.6--32.0 & 117 & $+0.71_{-0.88}^{+0.83}$ & $-0.30_{-0.81}^{+0.52}$ &  24.0   & 43.0 & 43.6 & 44.0 & 44.2\\
12 35 55.13 & $+$62 09 01.7 & 1.4--28.8 & 189 & $+0.10_{-0.38}^{+0.37}$ & $+1.29_{-0.41}^{+0.42}$ &  23.0   & 44.0 & 44.3 & 44.4 & 44.5\\
12 36 06.70 & $+$62 15 50.7 & 1.7--34.2 & 347 & $+1.89_{-0.27}^{+0.27}$ & $+2.07_{-0.38}^{+0.37}$ &  20.0   & 43.7 & 43.7 & 43.8 & 43.9\\
12 36 06.84 & $+$62 10 21.4 & 1.8--35.1 &  55 & \dots                   & $+0.10_{-0.52}^{+0.67}$ &  24.0   &$<$42.8 & 43.4 & 43.7 & 43.9\\
12 36 16.11 & $+$62 15 13.7 & 1.8--35.8 & 108 & $+0.37_{-0.55}^{+0.54}$ & $+1.34_{-0.52}^{+0.44}$ &  23.0   & 43.4 & 43.6 & 43.7 & 43.8\\
\\
12 36 22.66 & $+$62 16 29.8 & 1.7--34.7 &  71 & \dots                   & $-0.76_{-0.69}^{+0.61}$ &  24.2   &$<$42.7 & 43.6 & 44.0 & 44.3\\
12 36 29.11 & $+$62 10 45.9 & 1.0--20.1 & 178 & $+0.15_{-0.29}^{+0.29}$ & $+1.54_{-0.64}^{+0.74}$ &  23.0   & 42.8 & 43.0 & 43.2 & 43.2\\
12 36 32.59 & $+$62 07 59.8 & 1.5--29.9 & 136 & $-0.05_{-0.62}^{+0.59}$ & $+0.22_{-0.53}^{+0.51}$ &  23.8   & 43.1 & 43.6 & 43.9 & 44.0\\
12 36 35.58 & $+$62 14 24.1 & 1.5--30.1 & 217 & $-0.07_{-0.40}^{+0.40}$ & $+0.63_{-0.33}^{+0.33}$ &  23.7   & 43.3 & 43.7 & 44.0 & 44.1\\
12 36 36.75 & $+$62 11 56.0 & 0.8--15.6 & 650 & $+1.82_{-0.21}^{+0.22}$ & \dots                   &  20.0   & 42.4 & \dots & 42.7 & 42.7\\
\\
12 37 07.20 & $+$62 14 07.9 & 1.7--34.8 &  78 & $-0.52_{-0.81}^{+0.75}$ & $+0.74_{-0.49}^{+0.49}$ &  23.5   & 43.2 & 43.6 & 43.8 & 43.9\\
12 37 12.04 & $+$62 13 25.7 & 1.5--29.9 &  52 & \dots                   & $+0.04_{-0.71}^{+0.67}$ &  24.0   &$<$42.7 & 43.3 & 43.6 & 43.8\\
12 37 12.09 & $+$62 12 11.3 & 2.0--39.1 &  32 & $+1.34_{-1.81}^{+1.49}$ & $+0.23_{-1.30}^{+1.09}$ &  23.8   & 42.6 & 43.2 & 43.4 & 43.6\\
12 37 16.04 & $+$62 03 23.7 & 1.5--30.4 & 278 & $+1.14_{-0.29}^{+0.29}$ & $+1.63_{-0.37}^{+0.57}$ &  22.5   & 44.1 & 44.2 & 44.3 & 44.4\\
12 37 21.86 & $+$62 10 35.8 & 1.0--19.8 & 120 & $-0.63_{-0.24}^{+0.43}$ & $+0.79_{-0.68}^{+0.69}$ &  23.5   & 42.7 & 43.1 & 43.3 & 43.4\\
\enddata
\tablenotetext{a}{X-ray source position, taken from Table~1.} 
\tablenotetext{b}{Corresponding rest-frame energy (keV) for the observed 0.5--10~keV band.} 
\tablenotetext{c}{Total number of 0.5--10~keV counts used in the X-ray spectral analyses.} 
\tablenotetext{d}{Best-fit photon index for a simple power-law model (with Galactic absorption) in the indicated rest-frame energy bands; the uncertainties refer to the 90\% confidence level (for one interesting parameter).} 
\tablenotetext{e}{Estimated X-ray absorbing column density in logarithmic units of cm$^{-2}$. See \S3 for justification and column density analyses.} 
\tablenotetext{f}{Rest-frame X-ray luminosity, determined directly from {\sc xspec} using the best-fit photon indices. This has not been corrected for absorption.} 
\tablenotetext{g}{Rest-frame 0.5--8.0~keV luminosity in logarithmic units of erg~s$^{-1}$, determined from the rest-frame 5--20~keV luminosity and corrected for the estimated X-ray absorbing column density using our adopted AGN model; see \S3.5.} 
\tablenotetext{h}{Rest-frame 0.5--8.0~keV luminosity in logarithmic units of erg~s$^{-1}$, determined from the rest-frame 5--20~keV luminosity and corrected for the estimated X-ray absorbing column density using a simple absorbed power-law model; see \S3.5.} 
\end{deluxetable}
\clearpage
\end{landscape}

\end{document}